\documentclass[preprint]{aastex63}
\usepackage{rotating}
\usepackage{amssymb}
\usepackage{amsmath}
\usepackage{color}
\usepackage{lineno}

\accepted{January 30, 2022, in the ApJ}
\submitjournal{ApJ}
\shorttitle{Hot Halos in L* Galaxies}
\shortauthors{Bregman et al.}

\begin{document}

\title{Hot Extended Galaxy Halos Around Local L* Galaxies From Sunyaev-Zeldovich Measurements}


\correspondingauthor{Joel N. Bregman}
\email{jbregman@umich.edu}

\author[0000-0001-6276-9526]{Joel N. Bregman}
\affiliation{Department of Astronomy \\
University of Michigan \\
Ann Arbor, MI  48109  USA}

\author{Edmund Hodges-Kluck}
\affiliation{NASA/GSFC \\
Mail Code: 662 \\
Greenbelt , MD 20771 USA}

\author{Zhijie Qu}
\affiliation{Department of Astronomy \\
University of Michigan \\
Ann Arbor, MI  48109  USA}

\author{Cameron Pratt}
\affiliation{Department of Astronomy \\
University of Michigan \\
Ann Arbor, MI  48109  USA}

\author{Jiang-Tao Li}
\affiliation{Department of Astronomy \\
University of Michigan \\
Ann Arbor, MI  48109  USA}

\author{Yansong Yun}
\affiliation{Department of Astronomy \\
University of Michigan \\
Ann Arbor, MI  48109  USA}

\begin{abstract}

Most of the baryons in $L^*$ galaxies are unaccounted for and are predicted to lie in hot gaseous halos (T $\sim 10^{6.5}$ K) that may extend beyond $R_{200}$.  A hot gaseous halo will produce a thermal Sunyaev-Zeldovich signal that is proportional to the product of the gas mass and the mass-weighted temperature.  To best detect this signal, we used a Needlet Independent Linear Combination all-sky \textit{Planck} map that we produced from the most recent \textit{Planck} data release, also incorporating \textit{WMAP} data.  The sample is 12 L* spiral galaxies with distances of $3-10$ Mpc, which are spatially resolved so that contamination from the optical galaxy can be excluded.  One galaxy, NGC 891, has a particularly strong SZ signal, and when excluding it, the stack of 11 galaxies is detected at about $4\sigma$ (declining with radius) and is extended to at least 250 kpc ($\approx R_{200}$) at $> \,$99\% confidence.  The gas mass within a spherical volume to a radius of 250 kpc is $9.8 \pm 2.8 \times 10^{10}$ M$_{\odot}$, for $T_{avg} = 3 \times 10^6$ K.  This is about 30\% of the predicted baryon content of the average galaxy ($3.1 \times 10^{11}$ M$_{\odot}$), and about equal to the mass of stars, disk gas, and warm halo gas.  The remaining missing baryons ($\approx 1.4 \times 10^{11}$ M$_{\odot}$, 40-50\% of the total baryon content) are likely to be hot and extend to the $400-500$ kpc volume, if not beyond.  The result is higher than predictions, but within the uncertainties. 

\end{abstract}


\section{Introduction} \label{sec:intro}

One of the striking realizations of the past 30 years is that galaxies are missing most of their baryons \citep{mcgaugh10, dai2010}.  This comes about because we know the mass of the dark matter halo from the rotation curves of galaxies.  The rotation curves do not extend to the virial radius (we use R$_{200}$ henceforth, as it is similar to the virial radius), but a number of rotation curves extend to radii where dark matter dominates the potential.  The dark matter distribution is understood from models \citep{salucci19}, so we can determine the halo mass, M$_h$(R$_{200}$). The cosmic baryon fraction is 15.7\% \citep{Planck2018cosmo}, corresponding to a dark matter to baryon ratio of 5.37, so the associated baryonic mass is directly predicted. 

An accounting of the mass within the optical galaxy requires one measure the stellar mass of stars, plus corrections for no longer visible stellar remnants (e.g., \citealt{Loew2013}).  To this is added the mass of the various ISM phases, such as the HI, HII and H$_2$ disks of spirals, or the hot atmosphere for early-type galaxies. The ratio of the sum of these baryonic components to the predicted baryonic mass is $\sim 30$\% for an L* galaxy and can be as little as a few percent for less massive galaxies \citep{mcgaugh10}.  This constitutes the missing baryon problem for galaxies, which has motivated a number of observational and theoretical studies.

Theoretical models provide context for this problem and show that it is not just an accounting curiosity, but a crucial clue to understanding galaxy formation and evolution. The simplest model for the accretion of baryons were spherical in nature, which led to an accretion shock at large radius, producing a hot halo at $\sim$T$_{virial}$ that might contain significant mass \citep{mo2010}.  This picture is modified by radiative losses, leading to significant cooling in the inner region (50-100 kpc) while the outer part of the halo has a cooling time greater than the Hubble time \citep{BenO2018,hop18,nel18a,soko2018}.

These numerical simulations show that this model is modified in several major ways for galaxies near L* (e.g., \citealt{soko2018, Loch2020, Pand2021}).  The accretion of gas is generally highly asymmetric, flowing inward along low entropy streams that terminate within 0.1R$_{200}$.  The ensuing star formation leads to supernovae that heat their surroundings, driving winds from the galaxies, so some of the gas is rendered unbound and some remains bound as a hot Circumgalactic Medium (CGM; for spiral galaxies AGN heating is likely less important).  Models are in agreement that the CGM is turbulent and multi-temperature due to this feedback process.  However, models differ significantly as to the mass of the CGM, its temperature distribution, the fraction of baryons that lie beyond R$_{200}$, and the rate at which gas is cooling onto the disk of the galaxy (e.g., \citealt{Pand2020,Davies2020}).  The reason for these broad uncertainties is that the observational constraints on the hot gas properties of the CGM are limited.

Gas at T$_{virial}$ for a L* spiral has a temperature typically of $2-3 \times 10^6$ K, such as in the Milky Way \citep{henley12,henley13} or NGC 891 \citep{hodges2018}.  Some less massive galaxies, such as NGC 4631 have a lower temperature halo \citep{wang2001,yama2009}, as expected.  At these temperatures, a significant fraction of the thermal emission falls below the Galactic absorption threshold (0.2-0.3 keV), so these halos can be X-ray faint.  Also, the emission from external galaxies occurs in the energy band where the Galactic X-ray background is highest (mostly due to local hot gas processes), adding to the challenge of observing such objects.

Despite these challenges, diffuse X-ray emission is commonly detected around nearby edge-on L* spirals  \citep{Tull2006,LiWang2013}.
This X-ray gas is found to extend perpendicularly to the disk, and can be traced to about 5-10 kpc above the plane (e.g.,  \citealt{hodges2018}).  
Around the Milky Way, such hot gas is detected in both emission and absorption, where it 
is dominated by material within 50 kpc, about $0.2R_{200}$ \citep{miller15}.

Another approach to detecting the extended hot gas halo is through the thermal Sunyaev-Zel’dovich (SZ) effect\citep{Sunyaev80}, whereby the CMB photons are scattered by the hot electrons, thus distorting the spectrum.  Previous works have used ensembles of galaxies to identify a signal, using \textit{Planck} data.  The effort by the \cite{PlanckXI2013} and \cite{greco15} stacked $\sim 10^4$ galaxies per bin ($\Delta$logM$_* = 0.1$) and detected a signal down to about logM$_*$ = 11.3.  These are massive galaxies (generally early-type) and the stacking procedure sometimes includes a signal from the larger scale environment in which they are found (e.g., galaxy clusters).  Another stack of slightly less massive galaxies ($10^{12.6} \lesssim M_{500}$ M$_{\odot}$ $\lesssim 10^{13}$ M$_{\odot}$) by \cite{Singh2018} at typical redshifts of 0.1-0.14 combine X-ray and SZ data and infer that $\sim 20-30$\% of baryons in these galaxies are hot and produce the observed signal.  The total SZ signal for a single object, $Y$, is less than $2\sigma$ significance in most M$_*$ bins. 
The study of \cite{Ma2015} cross-correlated a galaxy sample with weak gravitational lensing \citep{Heymans12} with the \textit{Planck} data and obtain a signal in broad mass bins (i.e., $10^{12} - 10^{14}$ M$_\odot$) at about $3 \sigma$, with a signal extending beyond the virial radius.  
These studies do not provide information about the more ordinary L* spirals, nor do they address the emission region within R$_{200}$, issues that are addressed in this work.

The total SZ signal for a single object, \textit{Y}, is roughly proportional to $M_{gas} T_{virial}$, and when $M_{gas} \propto M_h$, simple scaling relationships show that the total SZ signal scales as $Y \propto M_h^{5/3}$, which is why rich galaxy clusters are most easily detected.  However, as \cite{Taylor03} point out, the signal that one detects depends on the angular size distance (D$_A$) in the usual D$_A ^{-2}$ manner and that nearby lower mass objects can also have a detectable signal.  The signal is measured against a background that is significant and reduces the S/N of the halos from nearby galaxies, which are extended at \textit{Planck} resolution ($10 \arcmin$; discussed below).  Despite this obstacle, we find that the SZ signal from nearby L* galaxies (D$_A <$ 10 Mpc) are at near-detectable levels, and when stacked together, reveal a clear signal to distances of about $R_{200}$ (250 kpc; we note that for our sample, D$_A$ equals the comoving radial distance, to high accuracy).  This nearby galaxy sample has a major advantage in that the SZ signal can be resolved at the resolution of \textit{Planck}, allowing one to avoid contamination from the optical galaxy and providing spatial information about the extent of the hot medium.

\section{Sample Selection} \label{sec:sample}

We based our target selection on the predicted strength of the SZ signal, as given by the $M_h^{5/3} D_{A}^{-2}$ scaling relationship. 
We chose galaxies close enough that, at \textit{Planck} resolution, the optical galaxy can easily be distinguished from the extended SZ halo.  The warm dust in the optical galaxy can cause a signal that contaminates the SZ signal and must be separated. 
One can determine $M_h$ with a sufficient amount of dynamical data, but that does not exist uniformly for each of the target galaxies or can be too uncertain.  Instead, we adopted the stellar mass in the K band as a proxy for $M_h$ (K$_{tot}$ from 2MASS; \citealt{2MASS}).  The range in the K band luminosity, or equivalently, $M_*$, is modest, with a median of log$M_{*} = 10.83$ and where the quartile values are at 10.76 and 10.90 (0.07 difference with the median; Table 1).  The difference between the median and the maximum or minimum values is 0.19 in log$M_*$. For the K band luminosity, the difference between the highest and lowest quartile boundaries is 36\%.

The distance is an important parameter, but all galaxies have multiple distance measurements, so we adopted consensus values from the \textit{NASA Extragalactic Database}, as given in Table ~\ref{tab:12galaxies}.  The target selection was the 12 L* spiral galaxies with the largest values of $M_h^{5/3} D_{A}^{-2}$, for distances less than 10 Mpc but beyond the Local Group, and without significant dust contamination from the midplane (e.g., NGC 6946 is excluded, with a Galactic latitude of 11.7\arcdeg ).  
The angular resolution issue led us to choose a maximum distance of 10 Mpc, at which the resolution of the maps, 10$\arcmin$, is 29 kpc.  This allows us to exclude the galaxy, which can have contaminating dust emission, while still having enough area to search for a signal from an extended halo.  

\begin{deluxetable*}{cccrrrrcccc}
\tablenum{1}
\tablecaption{Sample Galaxies: Basic Properties\label{tab:12galaxies}}
\tablewidth{0pt}
\tablehead{
\colhead{No.} & \colhead{Galaxy} & \colhead{Alt Name} & \colhead{RA} & \colhead{DEC} & \colhead{Gal \textit{b}} & \colhead{Maj Diam} & \colhead{Dist} & \colhead{K$_{tot}$}   & \colhead{log$M_*$}  & \colhead{Scaling} \\ 
    &          &          & \colhead{(degrees)} & \colhead{(degrees)} & \colhead{(degrees)} & \colhead{(arcmin)} & \colhead{(Mpc)} &        & \colhead{(M$_{\odot}$)} &     
}
\decimalcolnumbers
\startdata	
1   & NGC  253  &          & 11.8929   & -25.2922  & -87.97    & 37.20    & 3.22  & -23.75 & 10.81  & 14.85   \\ \hline
2   & NGC  891  &          & 35.6367   & 42.3467   & -17.42    & 13.18    & 9.52  & -23.92 & 10.88  & 14.02   \\ \hline
3   & NGC 1291 &          & 49.3275   & -41.1081  & -57.04    & 14.45    & 8.60  & -24.02 & 10.92  & 14.18   \\ \hline
4   & NGC 3031 & M81      & 148.8882  & 69.0653   & 40.90     & 28.20    & 3.67  & -23.97 & 10.90  & 14.88   \\ \hline
5   & NGC 3627 & M66      & 170.0624  & 12.9915   & 64.42     & 9.10     & 9.59  & -23.98 & 10.90  & 14.06   \\ \hline
6   & NGC 4258 & M106     & 184.7396  & 47.3040   & 68.84     & 18.60    & 7.28  & -23.83 & 10.84  & 14.19   \\ \hline
7   & NGC 4736 & M94      & 192.7233  & 41.1194   & 76.01     & 15.14    & 5.11  & -23.41 & 10.67  & 14.22   \\ \hline
8   & NGC 4826 & M64      & 194.1818  & 21.6830   & 84.42     & 13.80    & 5.40  & -23.29 & 10.62  & 14.09   \\ \hline
9   & NGC 5055 & M63      & 198.9554  & 42.0292   & 74.29     & 16.22    & 7.72  & -23.78 & 10.82  & 14.11   \\ \hline
10  & NGC 5194 & M51      & 202.4696  & 47.2344   & 68.52     & 15.85    & 7.19  & -23.64 & 10.76  & 14.08   \\ \hline
11  & NGC 5236 & M83      & 204.2540  & -29.8654  & 31.97     & 18.62    & 6.42  & -24.22 & 11.00  & 14.56   \\ \hline
12  & NGC 5457 & M101     & 210.8023  & 54.3490   & 59.77     & 30.20    & 6.86  & -23.63 & 10.76  & 14.11   \\ \hline
\enddata
\tablecomments{In the last column (11), Scaling, is, to a constant, log($M_*^{5/3} D^{-2}$), which was used in the sample selection process, along with the distance being closer than 10 Mpc.
}
\vspace{-0.5cm}
\end{deluxetable*}

\section{Map Selection} \label{sec:maps}

An all-sky map for the thermal SZ effect can be constructed from the nine \textit{Planck} channel maps, in concert with extracting the CMB, which has a known spectrum \citep{PlanckSZcat2016}.  The thermal SZ effect produces a distortion in the CMB with a known spectral shape but where the amplitude and angular size of the distortion is object-dependent.  This distortion causes a negative signal for $\nu <$ 217 GHz and a positive signal for $\nu >$ 217 GHz, and it is for this reason that there is a largely SZ-free band centered at 217 GHz.  The noise increases at low frequencies due to Galactic free-free emission, synchrotron emission, and spinning dust emission.  At high frequencies, contamination by dust and detector properties lead to larger noise.  Therefore, much of the weight of the SZ signal is given by the 100-353 GHz bands.  

From the \textit{Planck} data sets, thermal SZ maps, covering most of the sky were produced and made public. 
The Needlet Independent Linear Combination \textit{y} map (NILC; \citealt{Delab09}) uses the Internal Linear Combination (ILC) of maps to fit for the CMB plus other features, which can include the thermal SZ and contamination by dust or radio sources.  To this, 10 needlet function (window functions) are added, which are filters in multipole space to capture (and exclude) most of the contamination on various angular scales.  The maps are smoothed to the same $10 \arcmin$ resolution and they use the full set of high frequency maps (100 - 857 GHz) and the lower frequency data (30-70 GHz) but only for the large scale multipole space. 

These maps were constructed so that the SZ signal is net positive relative to the local background and that one can treat the NILC product as an image for the purpose of measuring surface brightness. For our initial studies, we used this public map rather than the public map produced by the Modified Internal Linear Combination Algorithm (MILCA; \citealt{Hurier13}), which, in our tests, had significantly higher excess kurtosis (i.e., longer winged tails in the pixel distribution), leading to more outliers and possible signal contamination.

These two public SZ maps were constructed from the Planck Data Release 2 (PR2) \citep{PlanckPR2}.  However, there were significant improvements in the processing of the Planck data in the two subsequent data releases, Data Release 3 (PR3) \citep{PlanckPR3} and Data Release 4 (PR4) \citep{PlanckPR4}.  For PR3, the improvements include better calibration and pipeline processing, along with changes in the map-making algorithms.  
However, PR3 contain striping residuals, which arise in an effort to remove the thermal fluctuations and far sidelobes as the satellite scans the sky.
Striping residuals were largely eliminated in PR4, which had less systematic noise on all angular scales when compared to previous releases.

We used the PR4 data, along with the WMAP 9 year data \citep{WMAP9yr} to construct SZ maps along the lines of the NILC approach, which is more fully discussed in Pratt et al. (2021).  There are a number of choices that need to be made for the needlet window functions and Gaussian weights, so there is not a unique set of choices for all purposes. For example, in the public NILC PR2 map, features on the scale of a few degrees was suppressed below its true value, due to the ILC bias and striping.
One cannot account for the ILC bias in the public maps due to limitations in the information provided regarding the window functions.
In the maps described by Pratt et al. (2021), the SZ signals on the scales relevant to nearby galaxies are not suppressed, which was tested by comparing input and extracted signals (below). 

\section{Data Extraction and Contamination}

\begin{figure}[t]
\begin{center}
\includegraphics[width=0.36\textwidth]{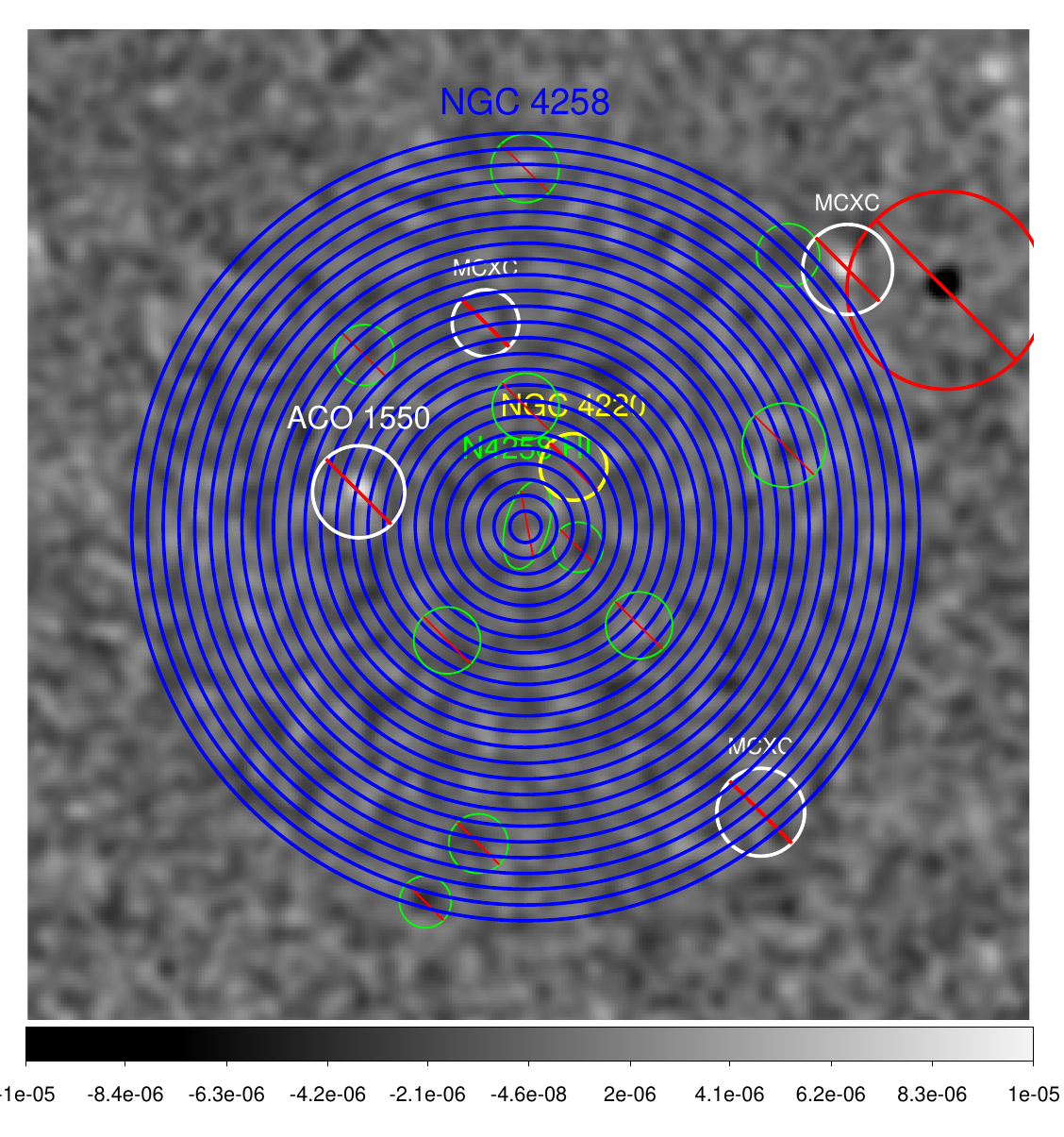}     
\includegraphics[height=0.3\textheight]{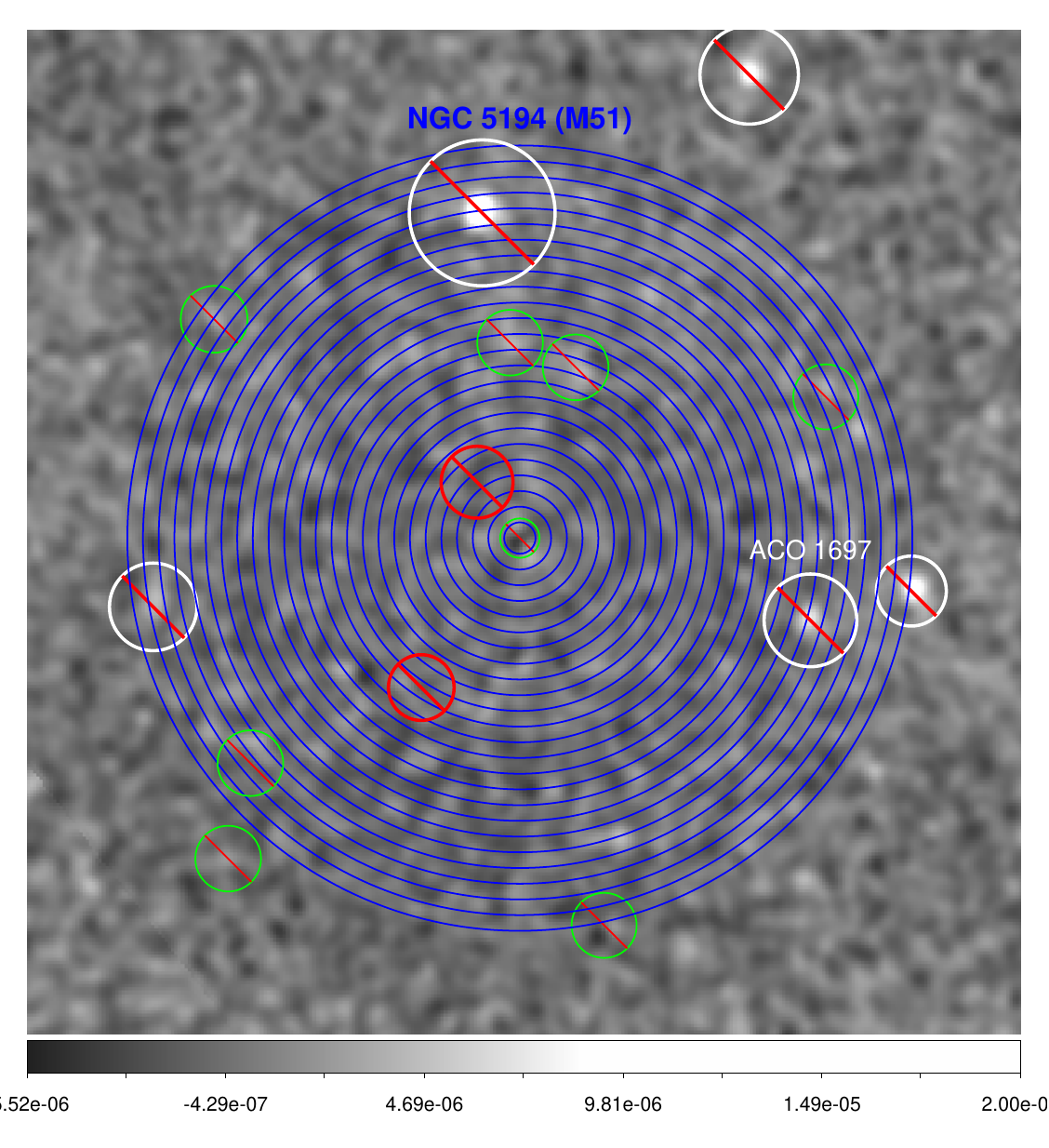}
\end{center}
\vspace{-0.4cm}
\caption{{\it Left panel}: The SZ map for NGC 4258 with annular regions and excluded regions, where the annular regions extend to a radius of 500 kpc (236$\arcmin$ ), with significant negative and positive regions excluded.  Among the excluded objects are Abell 1550 (white, E of center, z = 0.2814); NGC 4220 (yellow; NW; background galaxy, D = 20.3 Mpc) and some X-ray discovered clusters (MCXC designation; white).  Unlabeled positive brightness targets are possible cluster candidates and negative point sources are often AGNs. The strong radio source in red (W/NW of field, just beyond the edge of annuli) is 4C +49.22, a flat spectrum radio source at 0.33364, with a flux of 1-3 Jy in the Planck frequency bands. 
The elliptical HI region around NGC 4258 is also excluded ($13\arcmin \times 27\arcmin$ in semi-major axis). The greyscale range is from $-1 \times 10^{-5}$ (darkest) to $+1 \times 10^{-5}$.
}
\vspace{-0.15cm}
\caption{{\it Right panel}: The SZ map for NGC 5194 with annular regions and excluded regions as described in Fig. 1; the annular regions extend to a radius of 500 kpc (239$\arcmin$). Known QSOs and galaxy clusters are labeled (ACO 1697 is W of the center; z = 0.1808; the cluster N of center is Abell 1758, z = 0.28), while other bright excluded regions may be cluster candidates. The exclusion region around the optical galaxy (20$\arcmin$ diameter), also contains NGC 5195. The greyscale range is from $-5.5 \times 10^{-6}$ (darkest) to $+2 \times 10^{-5}$.
}
\vspace{-0.2cm}
\label{fig:N4258field}
\end{figure}

The process of constructing an all-sky SZ map leads to a map in which most regions away from the Galactic plane are close to a net zero value.  However, on an angular scale in which we extract an image (a few degrees), the local value of the background is statistically different than zero.  The mean background value can be either positive or negative, so we use a local background for signal extraction.  We exclude strong positive or negative sources that appear in the field and when we identify these sources, they are either clusters of galaxies (often in the \textit{Planck} catalog of SZ sources; \citealt{PlanckSZcat2016}), strongly IR-bright galaxies (easily identified in the 857 GHz maps), or radio-loud AGNs (also evident in the 100 GHz \textit{Planck} map, as well as in other catalogs).  When we identify the sources by eye, they are typically point sources with S\slash N $>$ 5. Automated source identification algorithms (e.g., \textit{SExtractor}; \citealt{Bertin96}) yielded nearly identical source lists.
This typically led to $\sim 10$ excluded regions per 50 square degrees, as seen in Figure ~\ref{fig:N4258field}.
Removing subsequently weaker sources  did not change the outcome of the SZ signal for a target galaxy.

The size of the exclusion regions around contaminating point sources were circles of 20\arcmin\ or 30\arcmin, depending on source brightness.  
Beyond these exclusion regions, the contamination by a point source to the surface brightness of the target galaxy was well below the signal being sought or the secular fluctuations in the background.
Also, within $R_{200}$ of a sample galaxy, we excluded other galaxies above 0.2L* (e.g., NGC 3628, near NGC 3627). 

Extended sources are rare but sometimes occur in our fields; these are nearby rich galaxy clusters that are in the \textit{Planck} SZ Catalog \citep{PlanckSZcat2016}, the Abell Cluster catalog \citep{Abell89}, or in a a meta-catalogue of X-ray detected clusters of galaxies (MCXC, \citealt{MCXC}).  In such cases, we would fit annular profiles to the SZ data to determine the radii at which the surface brightness contribution fell below that of the background fluctuations, which was typically at 1-2 Mpc.  
There were many cases where the exclusion radius could not be determined from the SZ data, due to signal weakness.  In this case, if $R_{200}$ was reliably determined from the X-ray data, we excluded the region within $R_{200}$. 
However, there were a number of clusters identified in the MCXC catalog with weak SZ signals and for which $R_{200}$ was not reliably estimated from the X-ray data, due to few X-ray source counts.  In those cases, we excluded a region of radius 1 Mpc.

Radio emission, especially from an AGN, can also contaminate the SZ map and almost always produce a negative signal.  These are not usually important within the galaxies in our sample, which are generally late-type galaxies with weak radio emission. 
We excluded all bright sources identified in the 143 GHz band and the 353 GHz band \citep{PlanckPCCS2}, many of which were similar. 
An example of a galaxy with strong radio contamination is the early-type system NGC 5128 (Cen A), which was not a candidate because it is not a spiral galaxy.
In addition, we excluded point like objects, positive or negative, that were not otherwise identified. These outliers were more than about $3 \sigma$ from the local field mean and are likely due to artifacts of the NILC method from minimizing the local covariance.   
Typical examples of  target regions, the annuli, and the exclusion regions are shown in Figure 1 and Figure 2. 

The mean SZ surface brightness values, $y$, were obtained in annuli defined in both angular and physical scales. When using angular scales, the width of a region was 10$\arcmin$, extending to at least 500 kpc ($4.1^{\circ}$ at D = 7 Mpc, the median of the sample), about $2R_{vir}$ for an L* galaxy.

\subsection{Excluding Contamination from the Optical Galaxy and Gaseous Disk}

We studied the contaminating effect on the SZ maps by dust emission from the galaxy and central point source emission. 
Low mass galaxies were chosen for this study as the SZ signal from a hot halo would be well below our detection threshold.  
One of the best cases is M33, with a dynamical mass about five times lower than our target galaxies \citep{Corb2014}, the anticipated net $Y$ signal would be about 14 times lower than for the L* galaxies in our sample, assuming that the SZ signal scales as $M_h^{5/3}$. 

\begin{figure*}
  \begin{center}
\includegraphics[width=4.4in]{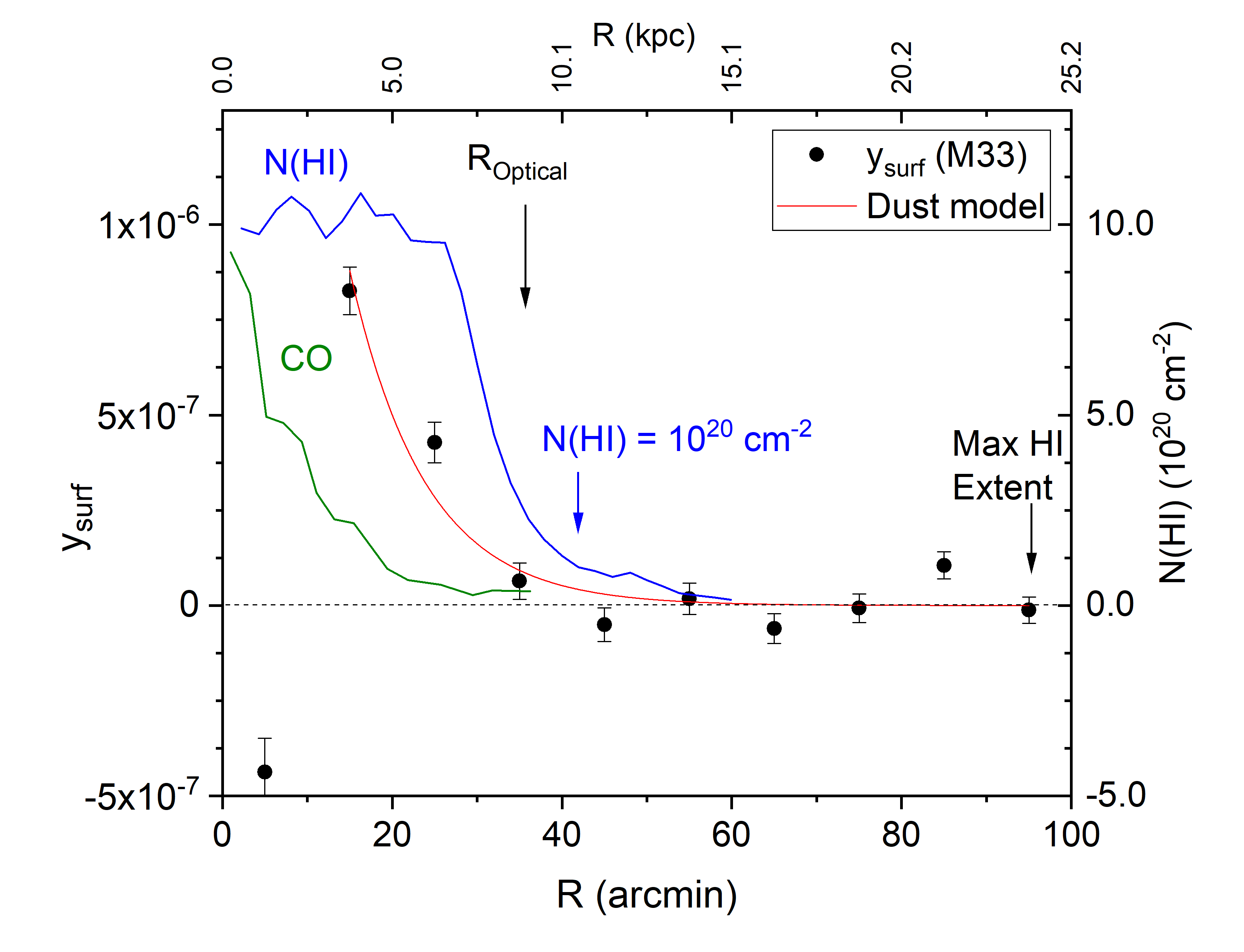}
  \end{center}
  \vskip -0.2in
\caption{
The SZ surface brightness measurement from the NILC map vs projected radius shows a dust signal interior to $30 \arcmin$ (7.6 kpc) of the center of M33.  The change in the SZ surface brightness $y$ from negative (first point) to positive (second, third points) is likely due to a decline in the dust temperature plus data processing procedures in map making. 
The optical radius marker is along the major axis and is the same value for the POSS1 103a-O \citep{Nilson73} and the RC3 D$_0$ \cite{deVauc95}; $R_{vir} \approx 120$ kpc.  The HI column density distribution (blue line; scale on right\cite{Kam17}) is more extended than the CO surface brightness (green line; scaled to fit the figure \citep{druard14}), which is typical of spirals.  Based on the NILC $y$ values, there is no measurable contamination to the null SZ surface brightness beyond N(HI) $= 1 \times 10^{20}$ cm$^{-2}$, so we exclude those interior regions in our galaxy sample.  A model for dust emission in spiral galaxies (red line;  \citealt{Smith2016}), fitted to the NILC $y$ values $R \geq 15^{\prime}$, also shows a sharp radial decline. 
The SZ signal from a gaseous halo around M33 is undetectable and is predicted to be $1.5\times 10^{-8}$ at R $= 55^{\prime}$. 
}
   \label{fig:M33nilc}
\end{figure*}

Differences in the temperature and size distributions of the dust lead to different spectral energy distributions.  Depending on these properties, dust emission can cause either a negative or positive signal in the SZ map.
This is because the SZ map is a linear combination of maps while the shape of the dust emission varies as a function of temperature, particle size distribution, and composition. 
We found that when N(HI) $< 1 \times 10^{20}$ cm$^{-2}$, dust emission is quite weak, such as in M33 (Figure ~\ref{fig:M33nilc}). This figure shows the dust contamination of the SZ signal has become unimportant at $R_{optical}$ (8.8 kpc for the major axis), which is 15\% smaller than the radius at which N(HI) $< 1 \times 10^{20}$ cm$^{-2}$ (10.6 kpc).  
This is consistent with sharp declines in dust emission around spiral galaxies, such as NGC 891 \citep{Bocchio2016}, or the exponential decline in dust emission in the combined signal of the spiral galaxies in the Herschel Reference Survey \citep{Smith2016}; this exponential decline, is shown in Figure \ref{fig:M33nilc}.
We note that when N(HI) $< 1 \times 10^{20}$ cm$^{-2}$ in the outer parts of galaxies, star formation becomes particularly weak \citep{Elmergreen2017}, so thermal emission by dust becomes faint.

For the spirals in our sample, the radius at which N(HI) $= 1 \times 10^{20}$ cm$^{-2}$ occurs is typically 20 kpc, although we obtained individual values for all sample galaxies.  
For seven galaxies, the HI exclusion region lay within the central 10\arcmin\ region, which is always excluded in our SZ analysis; these include NGC 891, NGC 1291, NGC 3627, NGC 4736, NGC 4826, NGC 5055, and NGC 5194 (M51a and M51b are separated only by 4.4$^{\prime}$, so they fit within the central exclusion region).  Of the remaining five galaxies, the HI exclusion regions were a 22.3$^\prime$ circle for NGC 253, a $10\arcmin \times 18\arcmin$ elliptical region (semi-major axis) for NGC 4258, a $10\arcmin \times 15\arcmin$ elliptical region for NGC 5236 (M83), and a $18\arcmin \times 20\arcmin$ elliptical region for NGC 5457 (M101).  A more complicated set of exclusion regions were applied to NGC 3031 (M81), as HI extends irregularly around it.  All HI zero moment maps are available from public databases and articles \citep{Boom2005,Walter2008,Haan2008,deBlok2018}.

\subsection{Notes on Individual Galaxy Fields}

This sample comprises well-known systems with optical luminosities near L*.  Notes on individual galaxies, their environment, and the characteristics of the SZ maps are given below.  If no mention is made of the SZ maps, they are typical and presented no special issues.

\vspace{0.15cm}
\noindent 1. NGC 253: This is a Sbc/Sc starburst galaxy that is inclined and located in the center of the Sculptor Group; it is the nearest galaxy in the sample.  It is the largest field ($2R_{200} = 8.9^{\circ}$) and is normal, with no unusual residuals. 

\vspace{0.2cm}
\noindent 2. NGC 891: An edge-on spiral (Sb) that is similar to the Milky Way and lies in a modest group, whose main member is suggested to be NGC 1023, although NGC 891 has a slightly brighter K band luminosity; they are separated by more than 1 Mpc.  NGC 1023, at D = 10.88 Mpc, lies beyond the 10 Mpc limit and is not included in this survey. 

\vspace{0.2cm}
\noindent 3. NGC 1291: An early-type spiral (SBa) in a region of typical noise properties. 

\vspace{0.2cm}
\noindent 4. NGC 3031 (M 81): This is the second closest spiral, nearly at the same distance as NGC 253.  It is the most massive spiral (Sab) in the M 81 group, which contains the starburst galaxy M 82.  M 82 and other galaxies were excluded in the SZ extraction, as were the extended HI regions connecting several group galaxies.  Within the 500 kpc radius region (7.7$^{\circ}$), there are noticeable residuals, presumably from Galactic dust, which were excluded.  

\vspace{0.2cm}
\noindent 5. NGC 3627 (M 66): This SBb system lies in the eastern cluster of the Leo I Group of galaxies, a cluster known as the Leo Triplet, which includes NGC 3628, NGC 3623 (M 65), and some less luminous galaxies; M66 is the most luminous.  The other galaxies were excluded in the flux extraction, as were two galaxy clusters in the outer parts of the extraction region.

\vspace{0.2cm}
\noindent 6. NGC 4258 (M 106): This Sbc galaxy has a LINER nucleus and elevated star formation, although there is no unusual central residual in the SZ map.  It is the brightest galaxy in a group with about 20 members \citep{Thilker07}. 

\vspace{0.2cm}
\noindent 7. NGC 4736 (M 94): This Sab galaxy has a LINER nucleus and is the brightest galaxy in the Canes Venatici I Group, which has 13 members, whose velocity structure has a strong component of Hubble expansion, rather than being a relaxed group \citep{Kara03}.

\vspace{0.2cm}
\noindent 8. NGC 4826 (M 64): This is a Sb galaxy that has a LINER designation and with prominent dust features. It is the brightest member of a poor group \citep{Tully2008}.

\vspace{0.2cm}
\noindent 9. NGC 5055 (M 63): This Sbc galaxy is listed as a H II galaxy and with a LINER designation. It has only one significant nearby galaxy, UGC 8313 and may be associated with the M 51 group, lying in the outer parts (SW), with a separation of $\sim 1$ Mpc.  

\vspace{0.2cm}
\noindent 10. NGC 5194 (M 51): This grand design Sc spiral (with a LINER designation) lies in a small group with NGC 5195 and UGC 8331 and are both to the SW of the M 101 group with a projected separation of $\sim 1$ Mpc.  They may be part of the M 101 group.

\vspace{0.2cm}
\noindent 11. NGC 5236 (M 83): This Sbc galaxy is located in a loose group of at least seven significant galaxies and about 1.5 Mpc (in projection) to the north of NGC 5128.  It is often assigned as part of the NGC 5128 group.  In the SZ maps, there are a few background clusters and foreground residuals that were excluded.

\vspace{0.2cm}
\noindent 12. NGC 5457 (M 101): This Sc galaxy lies at the center of a significant group \citep{Giur2000} with more than ten members (not including the small M 51 group to the SW). 

None of our targets lie in known X-ray emitting galaxy groups. 

\begin{figure*}
  \begin{center}
	\includegraphics[width=7in]{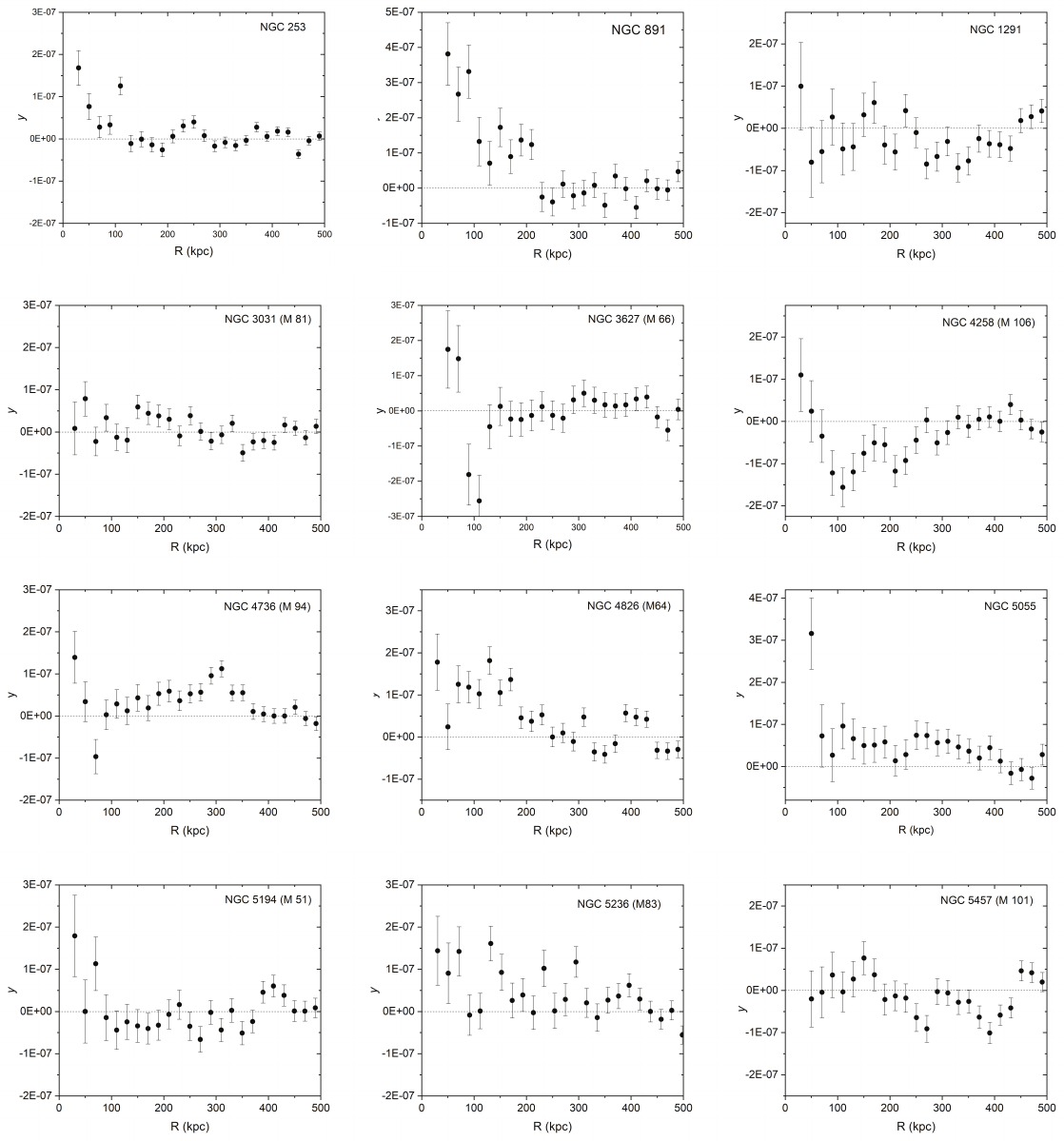}
  \end{center}	
\vskip -1.8in	
\caption{
The projected SZ surface brightness distribution \textit{y} (dimensionless) for the 12 target galaxies with errors.
The zero point of the background varies between target fields and here is defined as the mean flux in the 400-500 kpc region, adjusted to zero (dashed line).  The 1$\sigma$ errors shown here scale approximately as the inverse square root of the solid angle in an annulus, leading to statistical variations between galaxies. The vertical scales are the same ($5 \times 10^{-7}$) except for NGC 891 and NGC 3627, which is slightly larger ($6 \times 10^{-7}$).
}
   \label{fig:12galaxies}
\end{figure*}

\subsection{Uncertainties, Bias, and Recovery Tests}

These topics were discussed at length in section \S4 of \cite{Pratt2021}, so we provide a summary here, which parallels and makes use of that approach.  There are a few ways of estimating the uncertainty in the signal as a function of annular radius, but using the intrinsic instrumental uncertainties yields to errors that are unrealistically small.  This is because most of the noise comes from the uncertainties in separating the two stronger signals (CMB and dust) from the SZ signal, with most of the noise associated with the ``dust” map, which is actually the sum of all signals that are not the CMB or the SZ.  

One way of obtaining the uncertainty in annuli is from the dispersion in annuli between blank fields, which leads to a dependence of $\sigma_i \propto \Omega_i^{-0.4}$, where $\Omega$ is the solid angle of annulus $i$.  One might have expected a $\Omega_i^{-0.5}$ dependence, but the modest secular variations of the backgrounds in the SZ maps around the sky flattens the dependence on solid angle. 
The other approach is to perform a bootstrap analysis of multiple objects, such as the objects in our sample.  This leads to a similar but slightly larger set of uncertainties, so to be conservative, we used these errors. 

Another test involves the accuracy by which an injected signal can be recovered, with the central part excised, which contains the optical galaxy.  \cite{Pratt2021} used simulated galaxy groups, which are about the same angular size as our target galaxies, but with a stronger signal.
The input signal was put into the \textit{Planck} (Data Release 4) and \textit{WMAP} (Year 9) frequency maps, from which the NILC map is produced.  The recovered surface brightness, \textit{y}, as a function of radius is very close to the input signal (Fig. 4 in \citealt{Pratt2021}), with the difference being a modest bias term ($< 5\%$).  When recovering a stacked signal, there is little difference between the weighted average and the bootstrap procedure.

\begin{figure}
  \begin{center}
\includegraphics[width=4.5in]{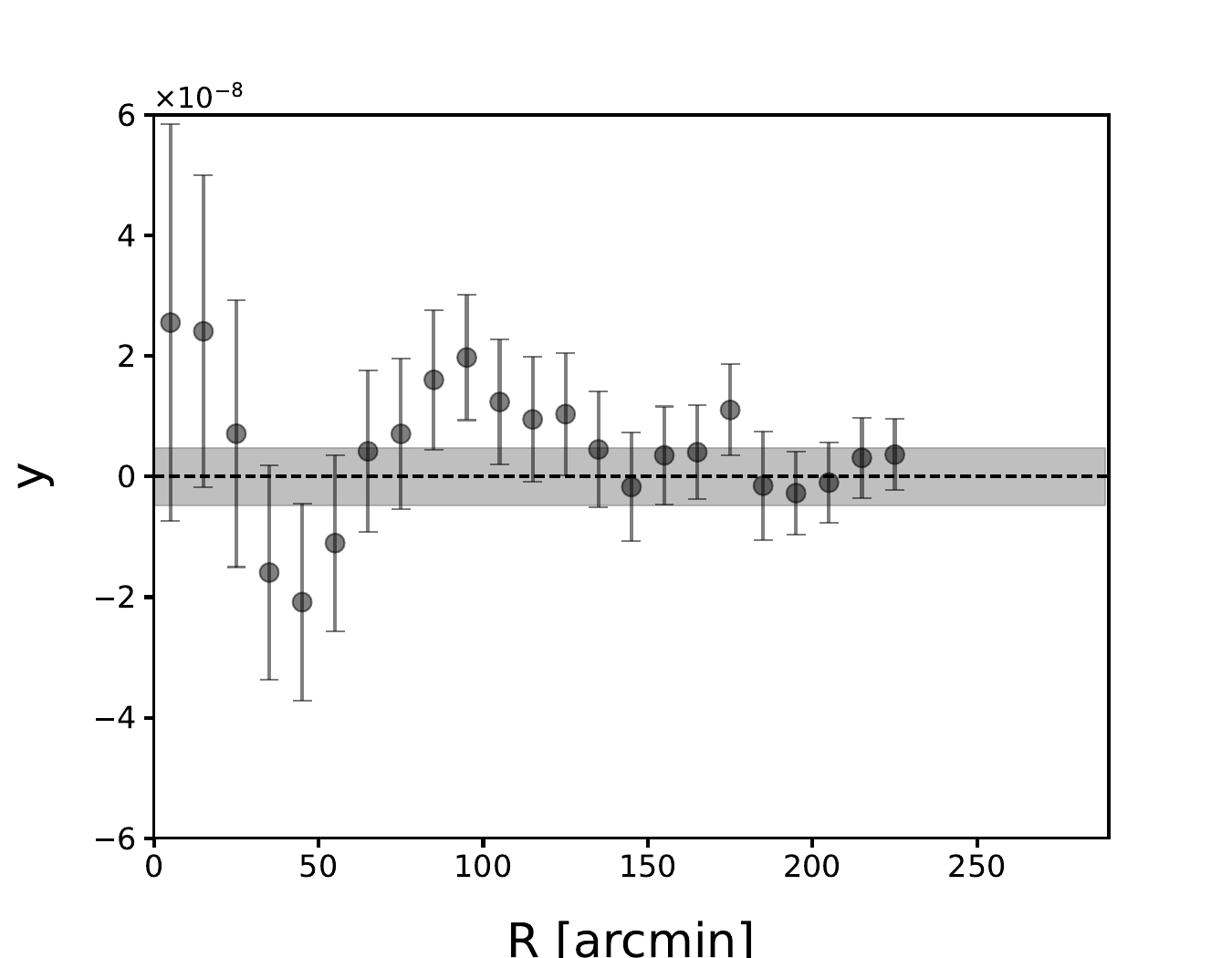}
  \end{center}
\caption{A stacked radial profile using 1000 random regions without known central sources and for 10$^\prime$ bins. The mean value determined by the outer 6 bins from $230^{\prime} - 290^{\prime}$ were subtracted off from each field as a local background subtraction. A constant value of $y=0$ is an acceptable fit, although a slight offset constant $y = 3.9 \times 10^{-9}$ is a better fit (the grey bar shows the $1 \sigma$ uncertainty. A fit with eq. (2) leads to a positive detection with S/N = 0.3 for $S_1$ ($1.02 \pm 0.30 \times 10^{-8}$)and an offset of $S_0 = 2.6 \times 10^{-9}$, so there is no significant signal intrinsic to random stacks (also, see Figure 3 of \citealt{Pratt2021}).
}
  \label{fig:DustCorrelation}
\end{figure}

One can use blank fields to determine if there are correlations with Galactic foregrounds. \cite{Pratt2021} provided a correlation analysis between Galactic dust and the SZ signal (their Figure 3). They show the 2-point correlation functions for the SZ background autocorrelation (i.e., excluding known SZ sources) and the cross-correlation between the 857 GHz map (a proxy for Galactic dust) and the SZ background. One finds a weak anti-correlation between the SZ background and Galactic dust, most prominent near 1$^{\circ}$. In order to see how such correlations may have affected our stacked results, we stacked the radial profiles using 1000 random fields shown in Figure ~\ref{fig:DustCorrelation}. One can see a slight ``dip then rise" part of the signal near $30^{\prime}-60^{\prime}$, which could explain the similar feature near 100 kpc in the stack of galaxies. However, these variations were not statistically significant, as the randomly stacked data were generally consistent with zero. These fluctuations, even if present, are unlikely to change the integrated value of $Y$(250 kpc) by more than 25\%.

\section{Results}

The goal of this program was to determine the SZ surface brightness quantity $y$, which is defined in the usual way as
\begin{equation}
y =\frac{\sigma _{T} k}{m_{e} c^{2}} \int _{-R_{max}}^{R_{max}}\, n_{e} (x) \, T (x) \, d x
\end{equation}
where $x$ is the line of sight coordinate and $R_{max}$ is the limit on the path length.  With an ideal detector, $R_{max}$ is the entire cosmic path length, including the monopole SZ signal.  However, we subtract a mean local background, removing a monopole term, and expect that any signal is local to the galaxy, where $R_{max}$ is several multiples of $R_{200}$, which should be constrained from radial surface brightness fitting. 
The total projected signal for a target, $Y$, is obtained by integrating $y$ radially. 
We begin by showing the stacked quantities for the entire sample and then identify that one galaxy is significantly brighter than the rest, so the stack is also constituted for the 11 galaxies without the outlier (NGC 891).  The radial shape of the SZ signal informs the radial dependence of the pressure and the extent of the signal.  This can be converted to a gas mass within the physically relevant radius $R_{200}$, upon adopting a mass-weighted temperature or temperature profile.

\begin{figure}
  \begin{center}
\includegraphics[width=4.5in]{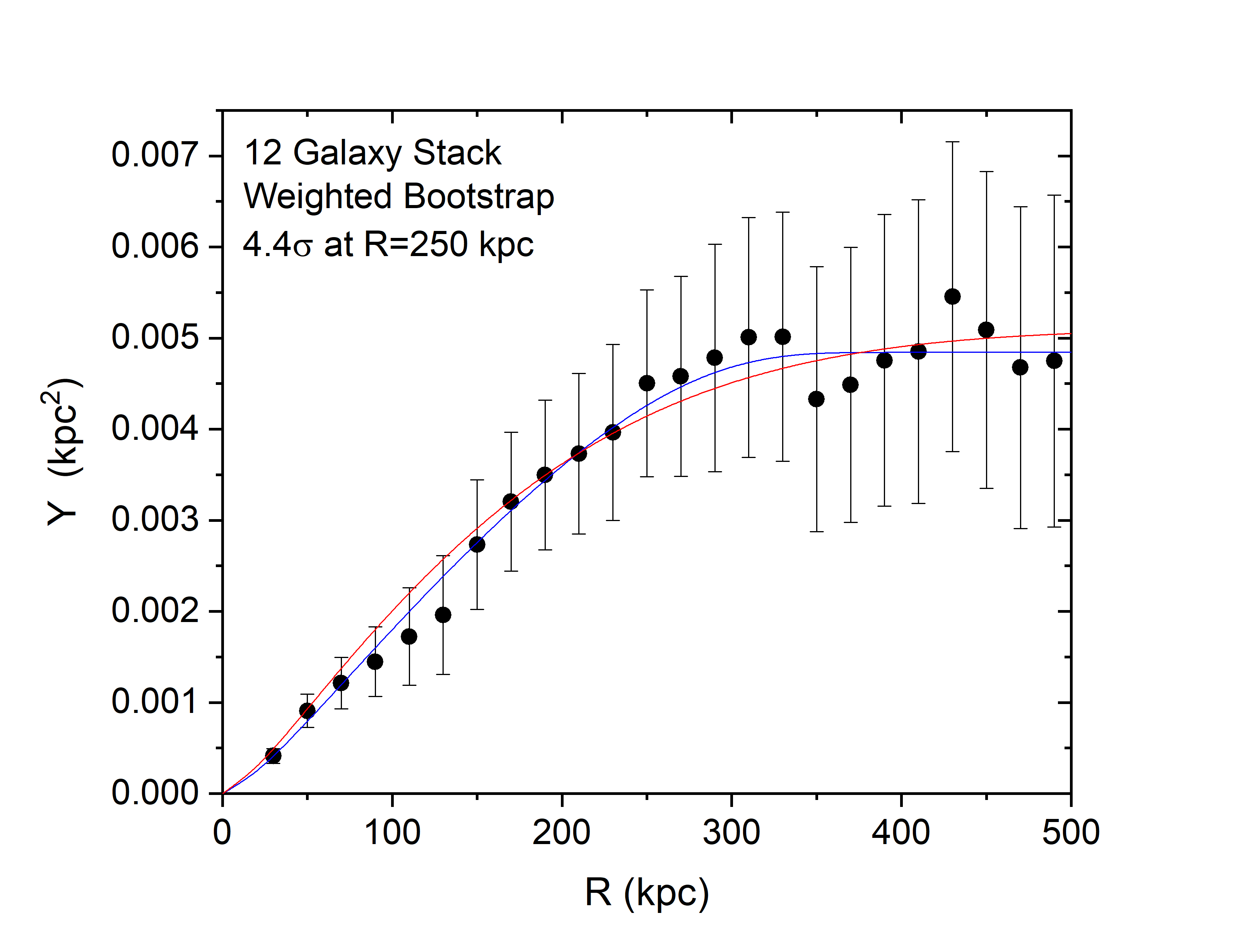}
  \end{center}
  \vskip -0.2in
\caption{
The integrated projected SZ signal ($Y)$ of the 12 galaxy sample, produced with the weighted bootstrap analysis, where the errors are the statistical uncertainties.  Using a $\beta$ model, a value of $\beta$ = 0.6 is preferred, while the two models only differ in the cutoff at large radius.  The red line has a Gaussian cutoff with $r_{cut} = 300$ kpc, while the blue line has a much sharper cutoff (see text), both being equally good fits. 
}
   \label{fig:Ysum12}
\end{figure}

Extraction of the signal from each galaxy was determined to the same outer radius, 500 kpc, which is approximately $2R_{200}$ (Figure \ref{fig:12galaxies}).  We stacked the galaxies in several different ways, both with and without weighting and obtained similar results.  The results shown here (Fig. \ref{fig:Ysum12}) are weighted by the scaling from Table 1, which is dominated by the inverse square of the distance ($D^{-2}$), as the range in masses is much smaller.  Nearly identical results are obtained for weightings entirely by $D^{-2}$ or by the statistical mean uncertainty in the signal.  The results of the weighted averages and the bootstrap procedure (using the same weights) also yielded nearly identical results and we present the weighted bootstrap results for $Y(r)$, as shown in Figure \ref{fig:Ysum12}.  We expected the 400-500 kpc region to have a null signal, as this region was used for the zero-point shift of the individual galaxy fields.  However, the signal appear to flatten by about 250-300 kpc, justifying the use of the 400-500 kpc region to define the background level.
An improved approach to determining the background is given below, where we adopt a parametric model that is guided by Figure \ref{fig:Ysum12}.

We fitted the data with a $\beta$ model with a smooth cutoff, defined as
\begin{equation} \label{beta_mod}
y(r) = S_1 (1+(r/r_c)^2)^{-3\beta/2+1/2} \times e^{(-(r/r_{cut})^{d})} + S_0 .
\end{equation}
One cutoff has $d=2$, a Gaussian, and with $r_{cut} = 300$ kpc, while the other has a sharper cutoff, with  $d=10$ and  $r_{cut} = 330$  kpc.
Despite the difference between these two cutoff terms, either provide an acceptable fit (shown in integrated form in Fig. \ref{fig:Ysum12}), indicating that there is too little information to determine the shape of the outer cutoff.  
Henceforth, we adopt $d=2$ and $r_{cut} = 300$ kpc for fitting surface brightness profiles.
Our fit to $y(r)$ leads to a best-fit value of $\beta = 0.61 \pm 0.06$, so we chose to fix 
$\beta = 0.6$ in much of our analysis, although we also examine the impact of other $\beta$ values in the gas mass determination.  This value of $\beta \approx 0.6$ can be due to the sum of the decline in the density and the temperature, an issue considered below. 

\begin{figure}
  \begin{center}
\includegraphics[width=4.5in]{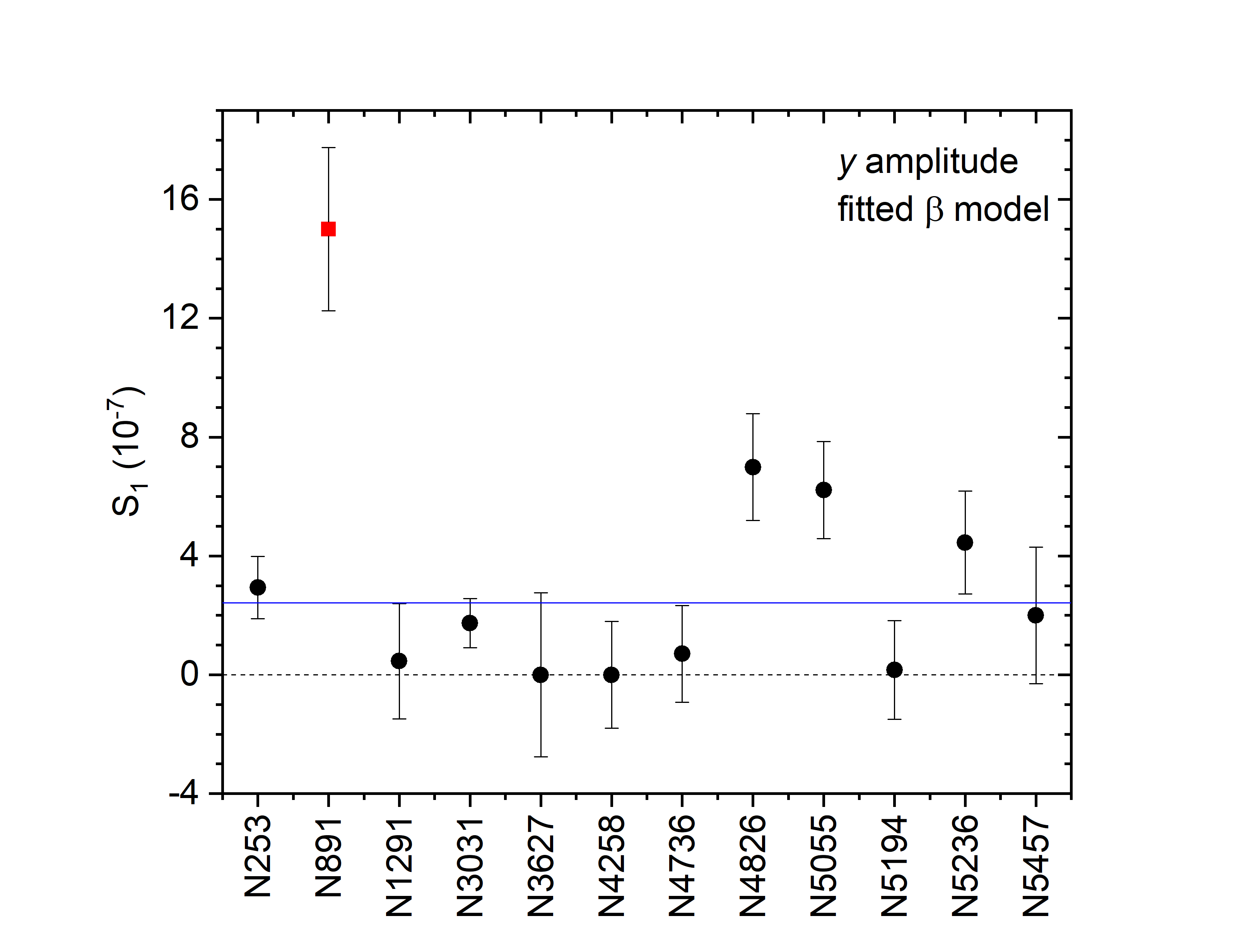}
  \end{center}
  \vskip -0.2in
\caption{The amplitude of the $y$ signal, based on the $\beta$ model as a function of NGC number.  The galaxy NGC 891 is an outlier, but the remaining 11 galaxies are consistent with a single population with a modest internal dispersion. For stacking, we use the 11 galaxies and examine NGC 891 separately.
The small negative amplitude (S$_1$) for NGC 3627 and NGC 4258 is set to zero here.
}
   \label{fig:S1vsNGC}
\end{figure}

The results from a stack are valid when the objects are drawn from the same population, without a large intrinsic dispersion about the mean.  We examine whether this is satisfied by fitting the projected $\beta$ models to $y$ for each galaxy, where we fix $\beta = 0.6$, $r_{core} = 10$ kpc, and adopt the Gaussian cutoff with $d = 2$, $r_{cut} = 300$  kpc.  
The result does not depend on the value of $r_{core}$, provided it is less than the inner excluded region (20-30 kpc), and for galaxies, the core radius is usually $< 10$ kpc (e.g., \citealt{miller15, goulding16}). 
There is a clear outlier in the SZ signal, where NGC 891 lies $4.6 \sigma$ from the median line defined by the weighted mean of the other 11 galaxies (Figure \ref{fig:S1vsNGC}).  

\begin{deluxetable}{rrcrr}
\tablenum{2}
\tablecaption{SZ Fitted Amplitude and S/N \label{tab:S1results}}
\tablewidth{0pt}
\tablehead{
\colhead{No.} & \colhead{Galaxy} & \colhead{Alt Name} & \colhead{$S_1$}  & \colhead{S/N}   \\
    &          &          & \colhead{(10$^{-7}$)} &       
}
\decimalcolnumbers
\startdata
1   & NGC 253  &          & 2.94       & 1.56  \\
2   & NGC 891  &          & 15.0       & 3.56  \\
3   & NGC 1291 &          & 0.46       & 0.13 \\
4   & NGC 3031 & M81      & 1.74       & 1.16 \\
5   & NGC 3627 & M66      & -2.08      & -0.42  \\
6   & NGC 4258 & M106     & -1.22       & -0.64  \\
7   & NGC 4736 & M94      & 0.70        & 0.24  \\
8   & NGC 4826 & M64      & 6.99        & 2.16  \\
9   & NGC 5055 & M63      & 6.22        & 2.12  \\
10  & NGC 5194 & M51      & 0.17        & 0.06  \\
11  & NGC 5236 & M83      & 4.45        & 1.43  \\
12  & NGC 5457 & M101     & 2.00       & 0.48   \\
\enddata
\tablecomments{The unphysical negative amplitudes for NGC 3627 and NGC 4258 were set to zero when determining the offsets that are needed for the weighted mean of the sample.}
\end{deluxetable}

The weighted mean of the 11 galaxies is $S_1 = 2.40 \pm 0.64 \times 10^{-7}$, where individual fits resulting in $S_1 < 0$ were set to zero.  This fit has a $\chi_{\nu}^2 = 2.08$, which corresponds to a chance occurrence of 2.5\%, suggesting that there is a modest intrinsic dispersion that has not been accounted for. The  $\chi_{\nu}^2$ can be reduced to lie within the 95\% confidence bounds by adding to each $S_1$ value an intrinsic dispersion of $0.6 \times 10^{-7}$, or 25\% of the weighted mean.  When including this intrinsic dispersion, the error weighted value becomes $S_1 = 2.43 \pm 0.67 \times 10^{-7}$.  This addition of an intrinsic dispersion is not large compared to the mean, so we conclude that, after excluding NGC 891, the remaining 11 galaxies can fairly be used for stacking. 

We note that two objects, NGC 3627 and NGC 4258, have negative fitted amplitudes (S$_1$).
While negative values are unphysical, these are within 1$\sigma$ of zero  (Table 2) and consistent with other near-zero values we measure. 
Improving the sensitivity to low $S_1$ systems requires more accurately accounting for contamination from foreground Galactic dust and understanding the noise contribution on multiple angular scales and from multiple frequency channels. Optimizing flat fielding and background estimation for SZ measurements of CGM will be the subject of future work (\S8.6). 

\subsection{Individual Galaxy Detection Significance}

We determined the significance of each galaxy in the sample in a few different ways, which included fitting a parametric function (above) as well as using the Pearson, Spearman, and Kendall tests to assess the presence of a correlation between $y$ and $r$.  Among the correlation tests, we choose to discuss the Kendall test, which is non-parametric, more robust against outliers, and usually yields a lower statistical significance, so it is the more conservative choice.  The six objects with the most significant correlations are NGC 253 (chance probability or p-value of 5.9\%), NGC 3031 (p = 9.2\%), NGC 5236 (p = 1.3\%), and three objects with p $<$ 0.2\%: NGC 891, NGC 4826, and NGC 5055.  This gives 50\% of the sample (six systems) with p $<$ 10\%, whereas 1.2 were expected.  There are three systems with p $<$ 0.2\%, where 0.024 were expected.  This indicates that the sample is inconsistent with a null signal and that a few galaxies were individually detected, although with significant uncertainties. 

The same galaxies with the smallest Kendall p-values have the highest significance in the $\beta$ model fits (Table 2), with some minor differences.  The confidence for the fit is highest for NGC 891 (chance probability is 0.02\%, or $3.56 \sigma$), followed by NGC 4826 (1.5\% or $2.16 \sigma$), which agrees with visual inspection.  These are followed by NGC 5055 (1.7\% or $2.12 \sigma$) and two galaxies with S/N $\approx 1.5$: NGC 253 (5.9\%); and NGC 5236 (7.6\%).  This parametric fitting leads to lower significances than the Kendall tests, due to fitting for two free parameters. From these parametric fits, 5/12 galaxies are significant above the 90\% confidence, whereas 1.2 would have been expected.  Three galaxies lie above the 98\% confidence, compared to 0.024 expected.

\subsection{Signal from the 11 Galaxy Stack}

\begin{figure}
  \begin{center}
\includegraphics[width=3.75in]{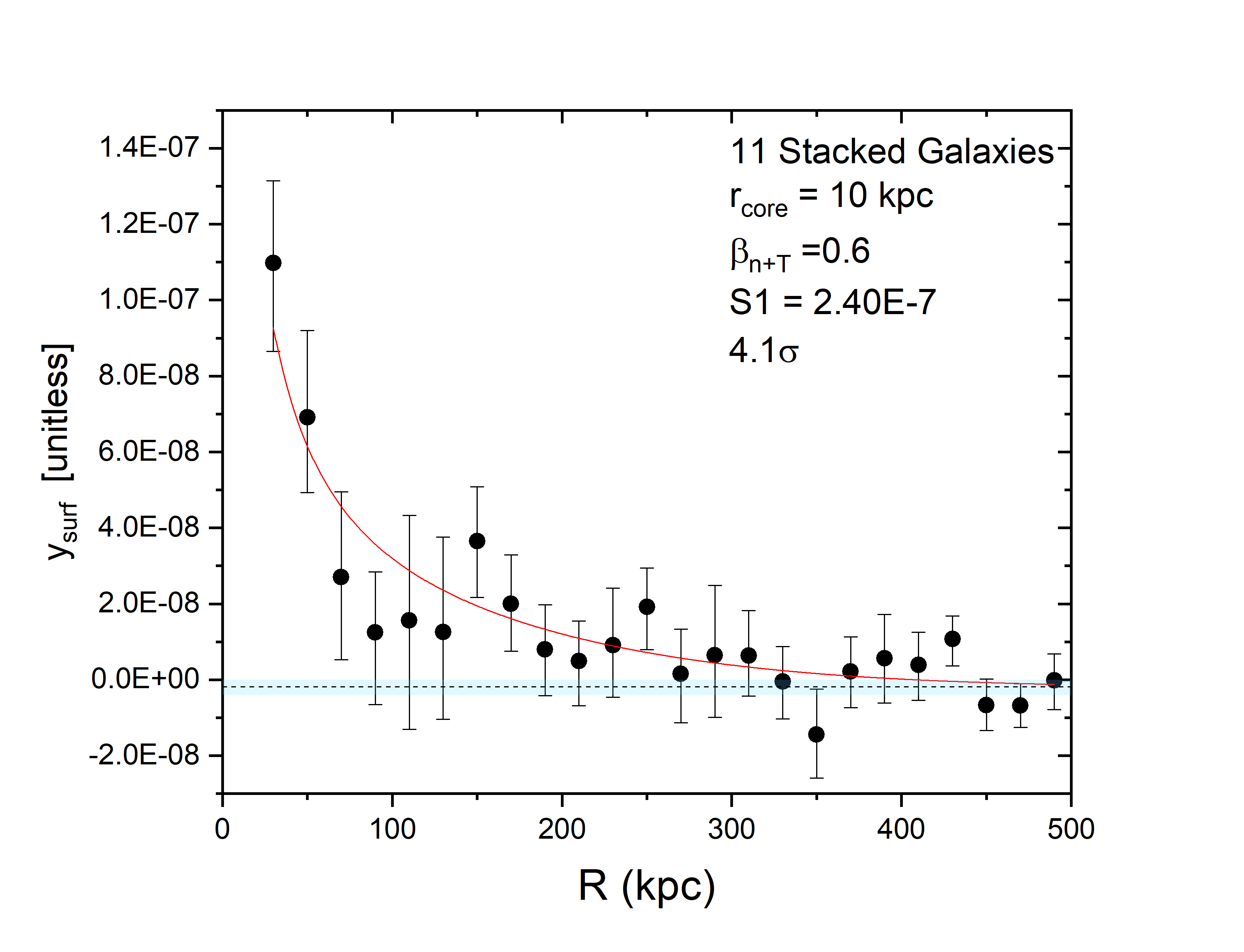} 
  \end{center}
  \vskip -0.2in
\caption{
The stacked signal from 11 galaxies (NGC 891 is excluded), obtained by the weighted bootstrap method and fitted with a $\beta$ model where $\beta = 0.6$ and a Gaussian cutoff with $r_{cut}$ = 300 kpc (red curve).  The zero point of the signal is given by the black dashed line and the $1\sigma$ error given by the blue band ($S_{0} = 0.18 \pm 2.08 \times 10^{-9}$). 
}
   \label{fig:y11gal}
\end{figure}

\begin{figure*}
  \begin{center}
\includegraphics[width=7.in]{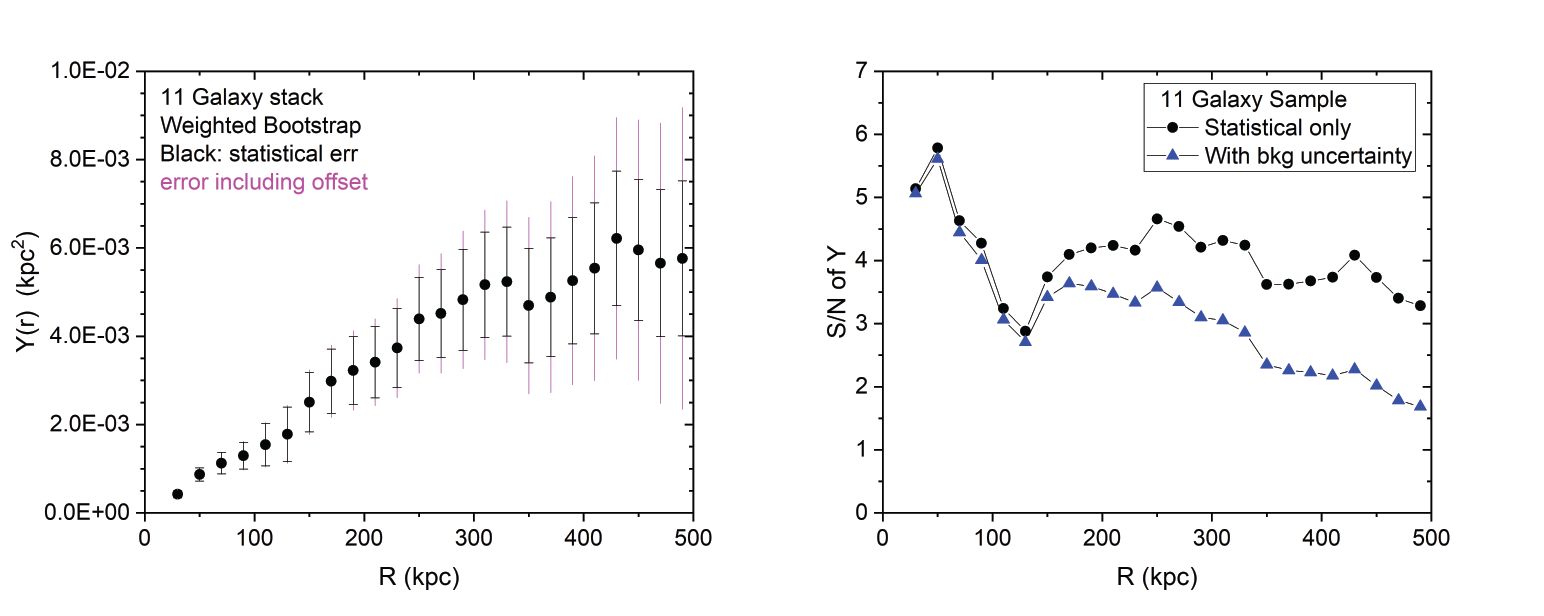}
  \end{center}
  \vskip -0.2in
\caption{
The integrated projected SZ signal, $Y(r)$ (left panel), based on the 11 galaxy sample (not including NGC 891) and produced with a weighted bootstrap analysis.  The black error bars show the statistical error while the total error (magenta) includes the zero point uncertainty errors.  The S/N of $Y(r)$ (right panel) is for statistical errors only (black circles) and the total error that includes the background uncertainty (blue triangles).
}
   \label{fig:Ytot11andSNR}
\end{figure*}

We calculated the weighted bootstrap results for both $y$ and $Y$, which are given in Figure \ref{fig:y11gal} and Figure \ref{fig:Ytot11andSNR}.  The significance of $Y$ requires no parametric model (so no negative data are excluded), while $y$ is fitted with the beta model, with the significance only weakly dependent on most model parameters, such as the nature of the cutoff at large radius. 

When inspecting the significance of $Y(r)$, we note that there are two uncertainties, the statistical uncertainty, and the uncertainty in the zero point offset of the stack, which is determined from the mean uncertainty of the summed individual extractions.  This uncertainty grows in importance with radius as $y$ becomes comparable to the offset uncertainty.  It may seem that the signal could become small at large radius, but it is actually telling us that the signal at large radius is poorly defined.  Physically, $Y$ cannot decrease with radius, as that would require a negative pressure, which is a physical impossibility.  At the minimum, one would expect $Y$ to become flat at large radii, if hot gas becomes greatly reduced beyond some radius, such as $R_{200}$.  There is likely an upper bound on $Y$, because if it continued to rise at a rapid rate, the inferred baryon fraction would exceed the cosmic baryon fraction.

For $Y(r)$, we show both the statistical uncertainty and the error including the uncertainty in the zero point offset.  For just the statistical uncertainty, the signal is more than 99\% significant through 500 kpc.  
When including the offset uncertainty, the signal is significant above the 99\% confidence out to 350 kpc, above the 98\% confidence out to 430 kpc, and above 95\% confidence to 500 kpc.  
At a projected radius of 250 kpc, the signal is $Y(r) = 4.39 \pm 0.93 (stat) \pm 0.79 (sys) \times 10^{-3}$ kpc$^2$; the sum of the two errors is $1.23 \times 10^{-3}$ kpc$^2$.  The presence of a signal is known better than the absolute value due to the offset in the background.  The signal at 250 kpc is 4.66$\sigma$ (statistical) and 3.57 $\sigma$ (total). 

Determining the integrated value at 500 kpc is uncertain due to the errors.  Relative to $Y(250 \, \text{kpc})$, the value at 500 kpc could be as much as a factor of two larger.  Using the last five data points (400-500 kpc), $Y(500 \, \text{kpc}) \approx 1.5 \, Y(250 \, \text{kpc})$, where the likely bounds are $Y(250 \, \text{kpc}) < Y(500 \, \text{kpc}) < 2Y(250 \, \text{kpc})$.

The $y(r)$ distribution has a S/N = 4.1, based on the same parametric fit applied to the individual galaxies.  The fit places constraints on the data and also determines the zero point offset independently. The normalization $S_1$ of the stack, $2.40 \times 10^{-7}$, is nearly identical to the weighted mean of the $S_1$ values from the individual 11 galaxies (discussed above), $2.42 \times 10^{-7}$.
The extent of the signal can be judged by eliminating inner bins until the confidence falls below some critical value.  We find that fits to $y$, using data in the 150-500 kpc range, still leads to a significant signal at the 99\% confidence level.  The significance falls below 90\% when only using data beyond 190 kpc.  There are still positive contributions beyond 190 kpc, as there is only one point that lies below the zero point (at 350 kpc) until reaching 450 kpc. 

\begin{figure}
  \begin{center}
\includegraphics[width=3.7in]{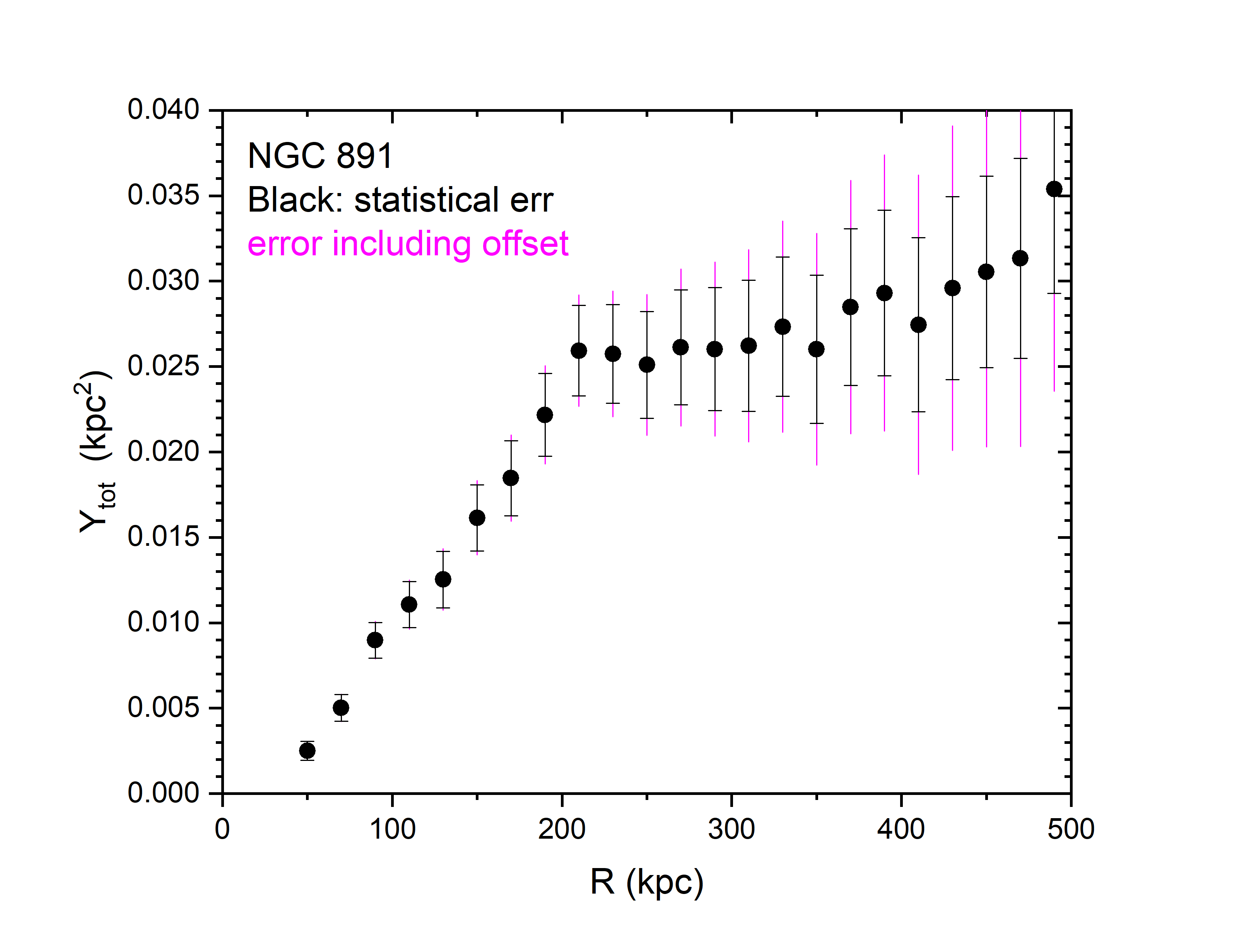}
  \end{center}
  \vskip -0.2in
\caption{
The integrated SZ signal, $Y(r)$ for NGC 891, where the errors and lines have the same meaning as the above figures.  The signal rises to 200 kpc and is then flattens at about 0.026 kpc$^2$, about six times larger than from the stack of the other 11 galaxies. 
}
   \label{fig:NGC891Ytot}
\end{figure}

We also consider the signals from the outlier system, NGC 891, where the normalization $S_1$ is six times larger than the stack of 11 galaxies.  The signal $Y(r)$ rises up steadily until 200 kpc, at which point it flattens to a value of $Y(r) \approx 2.6 \times 10^{-2}$ kpc$^2$, about 5 times larger than for the stack of 11 galaxies (Figure \ref{fig:NGC891Ytot}).  
At 250 kpc, the signal from NGC 891 is $Y = 2.51 \pm 0.31 (stat) \pm 0.27 (sys) \times 10^{-2}$ kpc$^2$; the sum of the two errors is $0.41 \times 10^{-2}$ kpc$^2$. 
The rise is steeper than for the stack of 11 galaxies and is more consistent with $\beta = 0.5$ (for $d = 2$ and $r_{cut} = 200$ kpc, $\beta = 0.53 \pm 0.10$). 

\section{Determination of Deprojected SZ Signal and Gas Mass}

Beginning with the projected SZ signal, we seek to determine the total SZ signal within a sphere of 250 kpc and also to place bounds on the total $Y$ signal within a sphere of 500 kpc.  
This deprojection requires knowledge of the three dimensional distribution of the pressure, which we will express by a power-law model with some sort of cutoff, which can be very sharp or a more gradual steepening.  
Most of the correction from a projected to spherical SZ signal occurs at radii greater than 250 kpc, where the SZ signal is most poorly known, so we calculate the correction for a variety of conditions. 

One limiting case, which leads to the largest correction, is where we assume a power-law density distribution for the pressure, $nT$, to 500 kpc (a $\beta$ model), with $\beta = 0.5, 0.6, 0.7$.  
For these three cases, the ratio of the SZ signal within a 250 kpc sphere to that projected on the sky is 0.66, 0.72, and 0.77.  Reducing the cutoff from 500 kpc to 400 kpc makes the ratios 0.68, 0.74 and 0.78.  Another model is where the signal cuts off more slowly, with a Gaussian cutoff in three-dimensional radius $R$, with a multiplicative term added to the power-law of the form $exp(-R^2/400^2)$, where $R$ is in kpc.  We note that this is different than the form used above, where the projected radius $r$ was used.  That cutoff, applied to the  $\beta = 0.5, 0.6, 0.7$ distributions, leads to ratios of 0.73, 0.78, and 0.82.  At fixed $\beta = 0.6$, decreasing this Gaussian cutoff radius from 400 kpc to 300 kpc raises the ratio to 0.89, while increasing the radius to 500 kpc, decreases the ratio to 0.74. 
The data seem to be most consistent with $\beta = 0.6$ and a Gaussian cutoff, although other models cannot be ruled out.  Here we adopt an intermediate ratio of $0.76 \pm 0.10$, where the $1\sigma$ range encompasses nearly all likely values. 
The observed projected signal within 250 kpc for the stack of 11 galaxies is $Y(r) = 4.39 \pm 1.23 \times 10^{-3}$ kpc$^2$, so the signal within a sphere of 250 kpc is $3.34 \pm 0.94 \times  10^{-3}$ kpc$^2$ (this includes the uncertainty in the conversion ratio). 

The relationship between $Y$ and gas mass is given by 
\begin{gather}
Y D_{A}^{2} =\frac{\sigma _{T} k}{m_{e} c^{2}} \int _{0}^{R_{200}}4 \pi  R^2 \, n_{e} (R) \, T (R) \, d R \  \\
= \ \frac{1.15 \sigma _{T} k}{m_{e} m_p c^{2}} M_{gas}(R_{200}) \, T_{avg}
\end{gather}

\noindent where $M_gas$ is the gas mass and $T_{avg}$ is the mass-weighted average.  We used a molecular weight such that $n_e = 1.15 n_p$, calculated from the abundances of \cite{Asplund2009}. 
The expression for $Y$ is  
\begin{equation}
Y = 2.36 \times 10^{ -3} \, \frac{T_{avg}}{3 \times 10^{6}\, \textrm{K}} \, \frac{M_{gas}}{10^{11} \textrm{M}_{\odot}} \, \frac{(7 \, \textrm{Mpc})^{2}}{D_{A}^2} \, \text{kpc}^2 . 
\end{equation}
The median distance for the galaxies in the sample is 7 Mpc, but we need to use the weighted distances, which leads to $D_{A}= 5.81$ Mpc, and using the above value of $Y$, which is been corrected for the conversion of the projected observation into a sphere, one obtains a gas mass in a sphere out to 250 kpc of 
\begin{equation}
M_{gas} = 0.98 \pm 0.28 \times 10^{11} \, \frac{Y}{3.34\times 10^{-3}} \frac{3 \times 10^{6}\, \textrm{K}} {T_{avg}} \ \textrm{M}_{\odot} \ . 
\end{equation}

The temperature of the hot halo for galaxies of this mass is known only close to the galaxy (e.g., within 10 kpc; e.g., \citealt{hodges2018}), and is unknown beyond about 50 kpc even in the Milky Way \citep{breg18}, so an accurate mass-weighted temperature is not available directly from X-ray data.  
We can estimate the temperatures from the virial mass, from X-ray observations, or from simulations. 
The total galaxy mass (M$_h$) within the virial radius is measured from dynamics or abundance matching, and the expected baryon mass is calculated by multiplying by the cosmic baryon fraction of 0.158 \citep{Planck2018cosmo}.  
For our sample, we used the abundance matching approach\citep{moster2010}, which gives $M_h \approx 2 \times 10^{12}$ M$_\odot$, and a total baryon mass of $\approx 3 \times 10^{11}$ M$_\odot$.  This $M_h$ is about 50-100\% larger than the value inferred for the Milky Way \citep{watkins2019, karukes2020}, and as the Milky Way hot halo gas is about $2 \times 10^6$ K \citep{breg18}, a value of about $3 \times 10^6$ K might be expected for our galaxy stack. 
Temperature profiles are determined in simulations, which yield a value near the virial temperature \citep{BenO2018,hop18,nel18a,soko2018}.  
However, the mass, temperature, and extent of simulated halos strongly depend on the uncertain feedback prescriptions \citep{Davies2020}. 
Here we adopt $T_{avg} = 3 \times 10^{6}\, \textrm{K}$ and hope that future work will lead to a more accurate treatment. 

\subsection{Halo Gas Mass of NGC 891}

The $Y$ signal around NGC 891 rises up sharply and turns over near 200 kpc (Figure \ref{fig:NGC891Ytot}), so foreground-background contamination from beyond 250 kpc is not expected to be large.  With either a sharp cutoff at 200 kpc, or a Gaussian decline of $exp(-R^2/200^2)$, the signal within a sphere of 250 kpc is within a few percent of the projected signal (for $\beta = 0.5$, which is a good fit to the data).  The signal of $Y(r) = 2.57 \pm 0.37 \times 10^{-2}$ kpc$^2$ and a distance of 9.52 Mpc, this corresponds to a mass of $M_{gas} = 2.01 \pm 0.29 \times 10^{12} \, (T/3 \times 10^6 \text{K})^{-1} $ M$_\odot$.  

NGC 891 was observed with long XMM-Newton and Chandra observations \citep{hodges2018}, where the X-ray emission is detected within 15 kpc of the plane. Near the star forming regions of the disk, the gas temperatures can be about 0.71 keV ($8.2 \times 10^6$ K), but at distances from the plane of 2-10 kpc, the temperature is closer to 0.2 keV ($2.3 \times 10^6$ K), with a significantly subsolar metallicity of $Z/Z_{\odot} \approx 0.14 $. The halo does not have an abnormally high gas temperature, so the above mass estimate should be correct.  
However, this implies that $M_{gas} \approx M_h$, which means that the baryon mass is about six times larger than the cosmic value if the galaxy contained all the baryons associated with the dark matter  (M$_h \sim 2 \times 10^{12}$  M$_\odot$, so M$_{baryon} \sim 3 \times 10^{12}$  M$_\odot$). This excess baryon mass is most likely associated with a galaxy group, although NGC 891 is not at the center of a rich group (see object notes above).  


\section{Contribution to the Baryon Budget}

We consider the major contributions to the baryon budget for the mean galaxy in the stack of 11 systems, being the stellar mass, the disk gas, the warm halo gas, and the hot ambient gas producing the SZ signal. The total baryon mass is $\approx 3 \times 10^{11}$ M$_\odot$, which is about a factor of three higher than the mass of the hot gas creating the SZ signal within 250 kpc ($0.98 \pm 0.28 \times 10^{11} \, (T_{avg}/3 \times 10^{6}\, \textrm{K}) \ \textrm{M}_{\odot}$).  The combination of stellar and disk gaseous mass is typical for an L* galaxy, is $\approx 6 \times 10^{10}$ M$_\odot$, which is 20\% of the cosmic baryon mass.  To this, we need to consider the mass of warm-hot gas in the halo, which is detected from UV absorption line measurements. 

A census of the warm ionized gas (T $< 10^5$ K) in the halo of the Milky Way \citep{Zheng2019,Qu2019MW1} and Andromeda \citep{Lehner2020} (similar in mass to our sample) leads to masses of $\sim 3 \times 10^{9}$ M$_\odot$, insufficient to close the baryon deficit \citep{breg18}.  
These masses do not make a significant contribution to the missing baryon issue. 

\begin{figure*}[ht]
  \begin{center}
	\includegraphics[width=12cm]{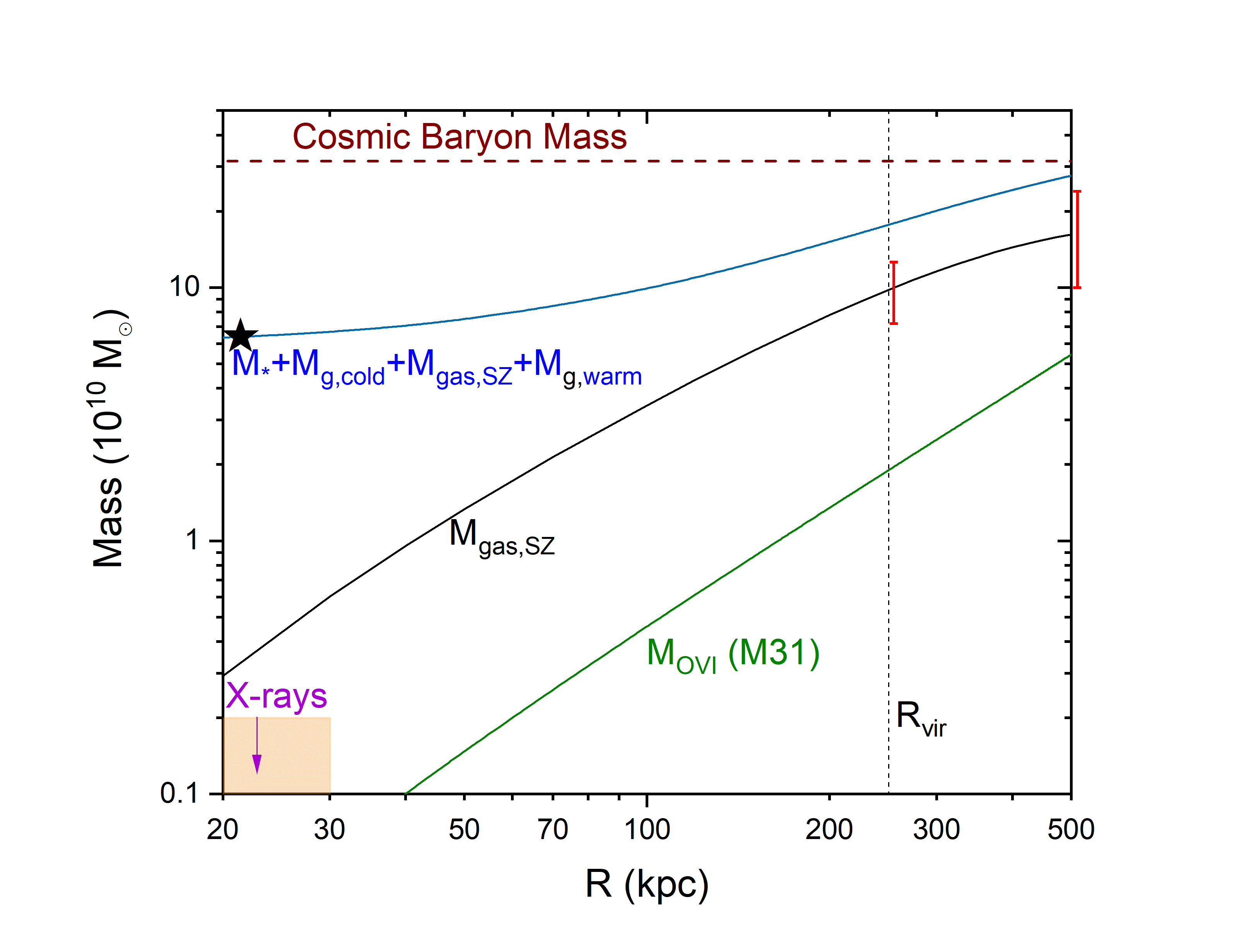}	
  \end{center}	
\vskip -0.2in
\caption{
The baryon masses of different components in a L* galaxy, including the stars (star symbol), cool disk gas, the hot gas that produces the SZ signal (black line), and the gas at T $\sim 10^{5.3}$ K responsible for the O VI absorption line gas (green; after \citealt{Lehner2020}); the sum is given by the blue line.  The uncertainties in the SZ gas are given at 250 kpc and 500 kpc by red vertical lines.  The sum of the stars, disk gas, hot gas from the SZ study, and the OVI bearing gas is shown as the dark blue line, which approaches the cosmic baryon mass at about 400--500 kpc. The brown box (lower left) shows the constraint that the hot gas from the SZ signal exceeds the X-ray limit \citep{Anderson2011,Anderson2016,jli18}.
}
\label{fig:Masses}
\end{figure*}

At higher temperatures, near $10^{5.5}$ K, gas may be traced by O VI.
In the Milky Way, this gas has an estimated halo mass component of $3 \times 10^9$ $M_{\odot}$  \citep{Qu2018a}.
However, it is about an order of magnitude higher in M31 \citep{Lehner2020}.
We wish to model these O VI columns as a function of radius, so we fit a model to the declining median projected N(O VI) with increasing radius $r$ of the form N(OVI) = $5.0\times 10^{14} \times (1+(r/100 \, \text{kpc})^2)^{-0.25}$. 
We note that this projected distribution, at large radius, would correspond to a three dimensional density that decreases as $R^{-3/2}$ (if fit with a $\beta = 0.5$ model).  
Our form for the O VI density is 
$n_{\rm OVI}(R) = 4.7 \times 10^{-12} \, (1+(R/100 \, \text{kpc})^2)^{3/4}$ cm$^{-3}$ . 
We note that the integral of the O VI mass diverges at large radii (500 kpc), increasing as $R^{3/2}$, so there must be an outer limiting radius. 
For calculating the gas mass contribution as a function of radius, we assume an ionization fraction of 0.20 for O VI and a metallicity of 0.3 Solar.  
We assume that this mass associated with O VI is typical of our stack of galaxies and show the sum of the different mass components in Figure \ref{fig:Masses}.

The contributions to the total expected baryon mass, given in Figure \ref{fig:Masses}, shows that the $M_{gas,SZ}$ equals the sum of the stars plus cold-warm gas at about 170 kpc, and becomes the dominant component at larger radii.  
At a radius of 250 kpc, the sum of the mass components constitutes about 50-60\% of the total expected baryon mass, which implies that a significant amount of mass lies beyond the virial radius.  The baryon deficit is nearly gone by 400-500 kpc, given the uncertainties in the measurements.  
It is possible that the O VI contribution is greater, if the metallicity or the ionization fraction is lower than assumed.  Without additional information, it is challenging to put uncertainties on the O VI contribution, so we do not include any. 

\section{Discussion and Conclusions}

This work showed that, with NILC SZ maps made from the most recent \textit{Planck} Data Release 4, we are able to see a clear SZ signal in a stack of 11 nearby ($D < 10$ Mpc) spiral galaxies. One galaxy was excluded from the original stack of 12 galaxies because it was about 6 times brighter than the stack, but the signals from the 11 galaxies do not suggest a large internal dispersion.  
This implies that the L* galaxies possess a substantial circumgalactic medium at an inferred temperature near the virial value. 
When converted to a gas mass at a constant mean $T_{avg} = 3 \times 10^6$ K, it equals the stellar mass plus cold disk gas at 170 kpc, it has $M_{gas}(250 kpc) = 0.98 \pm 0.28 \times 10^{11}$ M$_\odot$ and probably rises toward larger radii.
At a radius of 250 kpc, the sum of all baryons is about half of the cosmic value, so a significant amount of ``missing baryons'' lie at larger radii. 
In the following, we examine the relationship between this SZ signal and a variety of predictions and expectations. 

\subsection{Relationship to X-Ray Emission}

The combination of these SZ measurements with X-ray measurements further shows that some of the baryons must lie beyond the virial radius for spiral galaxies of similar mass (Fig. \ref{fig:Masses}).
The X-ray luminosity is proportional to the volume integrated emission measure, $\int_0^{R_{max}} \! 4\pi r^2 n_e^2 \mathrm{d}r$, so at fixed gas mass the emission measure rises for declining volumes.  
Diffuse X-ray emission around spiral galaxies is rarely detected beyond 20 kpc, which implies a limiting line of sight emission measure of about $\sim 1-3 \times 10^{-5}$ cm$^{-6}$ kpc, for typical \textit{Chandra} and \textit{XMM-Newton} observations \citep{anderson2010,Li2018, Bogdan2013, Bogdan2017}. 
Our fiducial SZ model is close to the X-ray detection limit at 20 kpc, and a somewhat lower metallicity (0.1 Solar) or a hotter temperature ($T > 3 \times 10^6$ K) would reduce the emission further.  
These surface brightness limits place a maximum value for the extended hot halo within 250 kpc of $\approx 1 \times 10^{11}$ M$_{\odot}$, for T $= 3 \times 10^6$ K and 0.3 Solar metallicity.
This excludes a hot halo within 250 kpc that contains all missing baryons. 

\subsection{Comparison to Other SZ Studies}

The extended nature of the hot halo is consistent with a study of more distant systems, where the \textit{Canada-France-Hawaii Telescope Lensing Survey} \citep{Heymans12} was cross-correlated with the \textit{Planck} SZ data \citep{Ma2015}, returning a signal at about $3 \sigma$.  
They divided the signal into two mass bins, the lower one being M$_h = 10^{12}-10^{14}$ M$_\odot$, and two radii bins of $0.01-1 \, R_{vir}$ and the other being $1-100 \, R_{vir}$.  Using both bins, they return a SZ signal at about $3 \sigma$.  Their data suggest that about 40\% of the SZ signal in this bin is extended beyond the virial radius, about a $2 \sigma$ result. This is consistent with our finding, although we use a narrow mass bin at M$_h = 10^{12.3}$ M$_\odot$.  Also, there is no discussion as to whether this extended emission is due to the low mass galaxy clusters that would lie in the mass range of their bin.

Previous investigations with \textit{Planck} data reported extended hot halos around stacks of significantly more massive galaxies \citep{PlanckXI2013, greco15,Singh2018,Pratt2021}, mostly elliptical systems about $\sim 10^2$ times more distant than our average galaxy.  
At masses of log(M$_*$/M$_{\odot}$) $\approx 11.3$, \cite{PlanckXI2013} found that $y$ is not point-like and appears to be larger than 2 Mpc in radius. This is much larger than $R_{200}$ for their galaxies and we suspect it is related to the contamination by the surrounding cluster.
The luminous early-type galaxies in their study are rare and often lie in rich environments, such as the center of a galaxy cluster.
In contrast, the L* spiral galaxies used here are much more common, occur in more ordinary environments, and define the critical “knee” of the galaxy luminosity function.

\subsection{SZ Scaling Relationships Relative to Galaxy Clusters and Massive Galaxies}

At large masses, such as galaxy clusters, the gas is nearly virialized value and nearly all of the baryons are contained within $R_{200}$ \citep{lagan2013} due to the depth of the potential well.  The thermal contribution due to feedback is small compared to the thermal energy from gravitational collapse, so under these conditions, the mass weighted temperature $T \propto M_h^{2/3}$.  This leads to a total signal $Y \propto M_{gas} \, T_{avg} \propto M_h^{5/3}$, so if an object has an SZ signal that lies on this relationship, one might conclude that the system contains all of its baryons.  

There is a problem when extending this conclusion to galaxies, as they are not self-similar to galaxy clusters.  
For example, the slopes of the X-ray surface brightness distributions are shallower in galaxies, with $\beta \approx 0.5$ \citep{goulding16}, than in rich clusters, where $\beta \approx 0.7-0.8$ \citep{Vikh2006}. 
This difference in $\beta$ is likely due to feedback having a greater effect on the temperature, which flattens the hydrostatic density distribution.  The gas temperature in spiral and elliptical galaxies is often $\sim$50\% higher than the virial temperature (e.g., \citealt{breg18}), whereas the rich cluster temperatures are close to the virial temperature and scale as $M^{2/3}$ (e.g., \citealt{Mantz2016} and references therein). 
Another difference between clusters and galaxies is that the fractional stellar mass, $M_* / M_h$, is typically 0.2, such as for this sample, while in rich clusters, a value of 0.05 is more typical \citep{lagan2013}.  

A comparison that can be made is from the scaling of the SZ signal from entire rich galaxy clusters, for which we can assume that 90\% of the cosmic value of baryons lie within a spherical volume bounded by $R_{200}$ \citep{lagan2013} and at the virial temperature.  
From the SZ signal from our galaxy stack, about 30\% of the baryons are in a hot halo within $R_{200}$, a factor of three below the value for rich clusters.
However, the hot gas temperature in galaxy halos is often above the virial temperature, so the net result is that one would expect $Y$ from our galaxy stack to lie about 50\% below the scaling relationship from rich clusters. 
That is approximately the result we find, although it is difficult to determine this with precision, given the scatter and normalization of the scaling relationship and the uncertainties in our stack $Y$ value. 
However, one can make more accurate and physically meaningful $Y -\text{M}_*$ relationships from detailed semi-analytic modeling and numerical simulation.

\begin{figure*}[ht]
  \begin{center}
	\includegraphics[width=13cm]{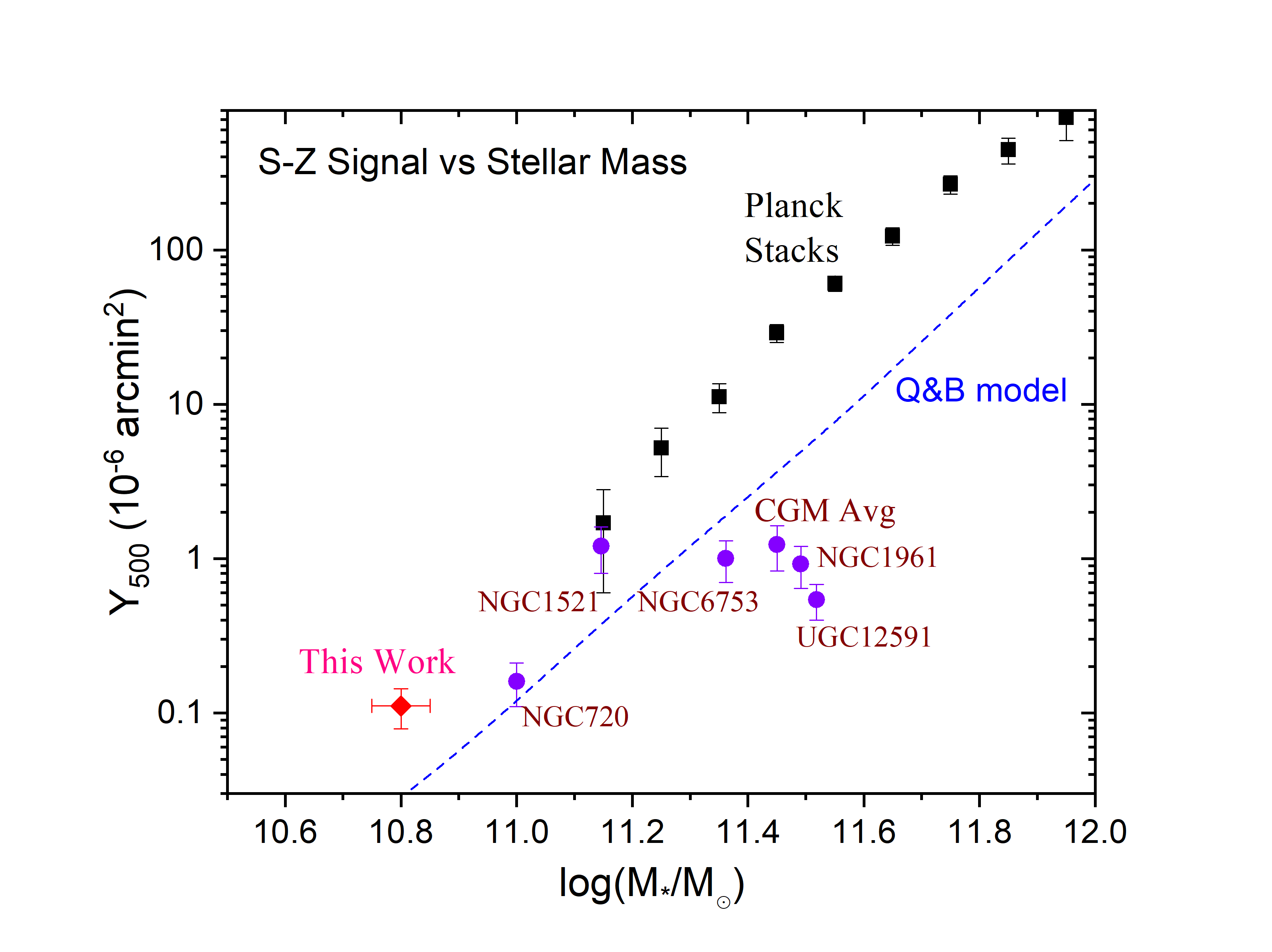}	
  \end{center}	
\vskip -0.2in
\caption{
The $Y(R_{500})$ signal vs the log of the stellar mass, where the red symbol is for the stack of nearby spirals in this work (placed at the common distance $D_A = 500$ Mpc). The purple circles are the SZ signal inferred from X-ray observations, mostly of massive spiral galaxies \citep{Li2018}, while the black squares are the stacked galaxies from \cite{PlanckXI2013}, mostly massive early type galaxies in the centers of galaxy groups and clusters.
The blue line is our scaling relationship for $Y$ for the hot halo gas mass of \cite{Qu2018a}. This relationship falls below our stacked value by a factor of three, it passes within a factor of two of the massive spirals, and lies an order of magnitude below the stacks from \cite{PlanckXI2013}.  
}
\label{fig:SZvsMstar500}
\end{figure*}

\subsection{Comparison to Semi-Analytic Models}

There have been a few semianalytical models for gaseous halos applied to the Milky Way and to galaxies of a range of masses \citep{faerman17,Qu2018a, Qu2018b,Voit2019,faerman19}.  These investigations include calculations of halo gas masses, SZ signals and entropy distributions.  This type of modeling requires that investigators make assumption for the temperature properties of the halo, as they are not yet sufficiently constrained by observations.  
For example, \cite{faerman19} model the Milky Way halo, where they assume an isentropic (constant entropy) halo, but with a ratio of specific heats $\gamma = 4/3$ to account for non-thermal pressure support (e.g., cosmic rays, magnetic fields).  The mass of their hot gaseous halo is $5.5 \times 10^{10}$ M$_{\odot}$, compared to the value from the stack, $9.8 \pm 2.8 \times 10^{10}$ M$_{\odot}$.  This is about a $1.5 \sigma$ difference, so not in significant disagreement, especially considering that the Milky Way galaxy is less massive than our stack of galaxies.  However, we note that the constant temperature profile assumed here is in tension with the temperature profile in \cite{faerman19}. 

Another work on hot halos is by \cite{Qu2018a}, who adopt a maximum temperature, such as 1-2T$_{vir}$, which does not change with radius.  This is not a true isothermal models because the gas radiatively cools, so there is a significant temperature distribution where the cooling time is less than a Hubble time; the mass-weighted temperature is not independent of radius.  The model was developed for comparison with the columns of ionic species, so the cooling function is corrected for photoionization (PIE) and here we use the results of this TPIE case.  
At R$_{200}$, the model gas mass is about half of the value for the stack, but the better comparison may be the SZ $Y$ value, which we have calculated here, based on the \cite{Qu2018a} work.  
This work provides M$_*$, M$_{gas}$ and mass-weighted temperatures as a function of M$_h$, and the  results are shown in Figure \ref{fig:SZvsMstar500}.  
This model line lies about a factor of three below the measured value from our stack, although it does pass within a factor of two of the massive spiral $Y$ values determined from X-ray data in Figure \ref{fig:SZvsMstar500}. 

It is not surprising that $Y$ from the stacks of \cite{PlanckXI2013} lie above our galaxy scaling, as the environments are extremely different. Most galaxies in their study lie in galaxy clusters and galaxy groups, which have hot gas with a mass that is often large compared to the central galaxy if it were relatively isolated.
This pressure of the cluster/group medium would likely raise the temperature and pressure of the gas within $R_{500}$ of the central galaxy, thereby biasing $Y_{500}$ to larger values relative to our relationship for non-cluster galaxies.  The magnitude of the offset would require extensive simulations and is beyond the scope of this paper. 

\cite{Voit2018} infer the CGM pressure profile around galaxies by assuming that the UV absorption line clouds \citep{werk14} are in pressure equilibrium with a hot ambient medium.  They find that this ambient pressure can be fit with a $R^{-1.8}$ profile, which is consistent with the pressure profile of our stack of galaxies. 

\subsection{Comparison to Predicted Signals from Large Scale Structure Simulations}

Hot gaseous halos are a generic feature of simulated galaxies \citep{BenO2018,hop18,nel18a,soko2018}. These simulations show that a fraction of the baryons form a stable rotating disk of stars and cool or warm gas in the central $0.1R_{vir}$. Much of the remaining gas forms a hot halo near the virial temperature that extends to near (or beyond) the virial radius, but the mass and extent of simulated halos strongly depend on the feedback prescriptions, which differ between EAGLE and \textit{IllustrisTNG}, such as in the AGN feedback algorithms \citep{Davies2020}. 

One simulation study, from the \textit{Feedback in Realistic Environments} program (\textit{FIRE}; \citealt{vandeVoort2016}) calculated the various SZ signals and baryon fractions at $z = 0 - 0.5$, within $R_{500}$, as a function of $M_{500}$.  We note that \textit{FIRE} includes stellar but not AGN feedback in this work.
Their hot baryon fraction within $R_{500}$ ($\approx 0.7R_{500}$, or 170 kpc for our sample) is similar to our value, $0.21 \pm 0.06$.
However, the more direct comparison is to their SZ parameter, where our signal lies toward the upper $1\sigma$ envelope of their simulation results. 

A study of halo masses with $EAGLE$ \citep{Davies2019} has a similar factor of two spread in the prediction of $Y$ and the gaseous halo baryon fraction at fixed $M_h$. 
Our hot halo mass (from the SZ signal) at $R_{200}$ is $0.31 \pm 0.09$ of the cosmic baryon fraction, indistinguishable from the $0.27$ value of the model. 
For $Y$, our value is about a factor of two above their median value at logM$_{200}$ = 12.3 but within the dispersion.  This could be explained if there is a temperature difference of a factor of two between the simulations and our data.

In a subsequent paper, \cite{Davies2020} considered the difference between \textit{EAGLE} and \textit{IllustrisTNG}, simulations that include feedback from both AGNs and stars. 
The outputs from both simulations are analyzed with the same approach, where the non-star forming fluid elements within $R_{200}$ are labeled as the CGM.
Their baryon fractions for the CGM differ by a factor of 2-3 for $11.5 < \log(M_{200}/M_{\odot}) < 12$ (\textit{IllustrisTNG} is higher), although at log($M_{200}/M_{\odot}) = 12.3$, they happen to be very similar, with baryon fractions for the CGM $f_{\rm CGM}/(\Omega_{\rm b}/\Omega_0) \approx 0.25$. 
These values are consistent with our stack value of $0.31 \pm 0.09$ at $R_{200}$. 
Overall, our halo gas properties and $Y$ value are in agreement with these three simulation codes. 

The signal beyond 250 kpc ($\approx R_{200}$) is poorly constrained by our data, so we focus on the signal within 250 kpc.  However, one can consider various contributions to the SZ signal at large radii and whether they would make a contribution near or within $R_{200}$.  One of the most important contributors could be from a hot intragroup medium.  None of the objects here have such a medium identified by extended X-ray emission. While the intragroup X-ray signal could be undetected, it may still make a SZ contribution.  We suggested that such a contribution occurs for NGC 891.  However, we do not know the distribution of faint, extended SZ signals associated with the other $L^*$ galaxies and obtaining such data will be very challenging.  Of comparable importance is the temperature distribution of the gas in our target galaxy, which is sensitive to feedback, such as from AGNs (e.g., \citealt{LeBrun2015}). 

One might consider a statistical approach, such as the contribution from the two-halo term.  
This has been discussed by a variety of authors \citep{Hill2013, Hill2018, Tanimura2020, Rotti2021, Pratt2021, Moster2021}, where most authors find it to be of minor importance in galaxy clusters, for multipole moments of $l \approx 10-30$ (for our mean galaxy, 2$R_{\rm 200}$ corresponds to about 5$^\circ$, or $l \approx 20$). 
The two-halo term can be important for analysis when the target is not resolved, an example of which occurs in  \cite{Tanimura2020}, who studied the SZ signal from luminous red galaxies at $z \approx 0.3$, where the HWHM of the beam (5$'$) corresponds to 1.5 Mpc, or about 2.2$R_{200}$ for their galaxies.  They find the signal extends beyond 5$'$, so they consider whether a two-halo term could be responsible for this very extended emission. 
There are various assumptions that go into this calculation, as the SZ signal has to be assigned to halos as a function of mass, yet it is not known a-priori in this mass range.  
Another uncertainty is the contribution from an extended hot gaseous group medium surrounding these luminous red galaxies. 
Putting aside these concerns for the moment, \cite{Tanimura2020} finds that a two-halo term can be the most important contributor at distances beyond 2.6$R_{200}$ (1.8 Mpc for their galaxies) and that this term is relatively flat, decreasing from $y = 3.6 \times 10^{-8}$ to $y = 2.5 \times 10^{-8}$ (30\%) from 0-10$'$ (0-3 Mpc).
The two-halo contribution from $L^*$ galaxies, with halo masses of $\sim 10^{12}$ M$_\odot$ was considered by \cite{Hill2018}, which requires one adopt pressure profiles, as they are not known for halo masses below $\sim 10^{13.5}$ $M_\odot$.  They predict that most of the signal is from the two-halo term and $y$ is fairly flat to 3 Mpc, with a similar 30\% decline and $y$ values. 

Our situation is quite different in that we resolve our target galaxies, which are the dominant halos within about 500 kpc of the centers of the target; our masking excludes lower mass galaxies, along with background objects.  It is unlikely that these surrounding lower mass halos contribute much to $y$, but if they did and it was similar to the results of \cite{Hill2018} and  \cite{Tanimura2020}, it would be manifest as a nearly flat elevated background.  The absolute level of the background cannot be measured accurately enough to test this prediction, due to the way the SZ maps are constructed.  However, if there were a nearly flat two-halo signal on these scales, it would be subtracted away by our background removal procedures.  Therefore, the two-halo component should contribute very little to our SZ signal. 

\subsection{Future Prospects}  \label{future}

The primary limitation in extracting a SZ signal for nearby galaxies (resolved by a 10$\arcmin$ beam) is not the sensitivity of the instrumentation -- it is the uncertainty introduced by the separation of the Galactic signal.  \cite{Pratt2021} improved upon the production of SZ maps by using the most recent \textit{Planck} data release and window functions that do not suppress the SZ signal on the scale of relevance to this study. We continue to develop better algorithms to address the separation of the dust contamination, using existing data. 
Further improvement should be possible by introducing more channels, such as the probe concept \textit{PICO}, which is designed to have 21 channels \citep{PICO2019}.  
Another improvement might result from higher angular resolution, as would be possible with such as CMB-S4 \citep{CMB-S4}.  Higher resolution could help to more accurately remove contaminating features.
It might be possible to improve results by combining the \textit{Planck} and \textit{WMAP} data with ground based programs, such as \textit{SPT} \citep{SPT2011} and \textit{ACT} \citep{ACT2016}, and we hope to explore that in the future.

Until new instruments are possible, we can expand the size of the stack of resolved galaxies by going to greater distances.  The disadvantage of using more distant galaxies is that a larger central exclusion region is required to avoid contamination by the optical galaxy signal. 
The goal would be to improve the signal in the $0.5-2R_{200}$ region so that we can understand the properties of the SZ signal in this outer region.  The practical distance limit is about 50 Mpc, where the 10$^\prime$ beam has a linear size of 145 kpc.

The temperature of the hot gas is needed to infer the gaseous mass and one possible way to obtain it is to bring together kinetic SZ measurements (kSZ) with the thermal SZ observations (tSZ).  The kSZ signal is about five times weaker than the tSZ signal, but its detection is reported for large samples of galaxy halos, using \textit{Atacama Cosmology Telescope} DR5 and \textit{Planck} data \citep{Amodeo2021,Schaan2021}.  \cite{Schaan2021} identifies galaxies where the mean virial halo masses are $3 \times 10^{13}$ $M_{\odot}$ ($z \approx 0.55$) and $5 \times 10^{13}$ $M_{\odot}$ ($z \approx 0.31$), which are more than an order of magnitude larger than the halo mass for our $L^*$ galaxies, $2 \times 10^{12}$ $M_{\odot}$.  They studied the properties of galaxy groups, not individual $L^*$ galaxies.  
\cite{Amodeo2021} study halos of similar mass, $3 \times 10^{13}$ $M_{\odot}$ ($z \approx 0.55$), so their temperatures ($1.3 \times 10^7$K in the center) and temperature profiles, which have large uncertainties, cannot be applied to our sample. 
Even with advances in instrumentation, it may be difficult to isolate the kSZ signal from galaxies of the low halo masses studied here. 

For the conversion of the SZ signal into a mass, the primary uncertainty is the temperature profile.  
The temperature profile of gas at $\sim10^{6.5}$ K can only be obtained from X-ray observations. 
Such halos should be detectable in absorption lines from the resonance lines of O VII, O VIII, C VI, and Ne IX, to name a few \citep{breg15}.  Future missions, such as \textit{Athena} \citep{barcons17}, \textit{Arcus} \citep{smith16}, or \textit{Lynx} \citep{gaskin18} will have the sensitivity to make major contributions in such absorption line studies.  Absorption lines from such hot halo gas should be detectable to at least $R_{200}$, and likely beyond.

Emission line studies with future instruments (calorimeters) will probe the hot halos as well, although to smaller radii, such as $\sim 0.5R_{200}$ for L* galaxies.  The primary missions that could survey galaxies to significant radii are \textit{Athena}, \textit{Lynx}, the \textit{Hot Universe Baryon Surveyor} (\textit{HUBS}) \cite{HUBS2020}, or \textit{Super DIOS} \citep{SuperDIOS}.  
These emission studies will give us temperature and metallicity profiles for individual galaxies. When combined with absorption lines, permit greater insight into the metallicity and the filling factors.
The X-ray observations, combined with SZ studies, will provide the definitive description of the hot halo gas.

\acknowledgments

We thank the many individuals who offered guidance and insight, including Mathieu Remazeilles, Renato Dupke, Chris Miller, Oleg Gnedin, Ben Oppenheimer, Mike Anderson, Jeff McMahon, and a very thoughtful referee.
We are grateful for support from NASA through the Astrophysics Data Analysis Program, awards NNX15AM93G and 80NSSC19K1013.
The work is based on observations obtained with \textit{Planck}, http://www.esa.int/Planck, an ESA science mission with instruments and contributions directly funded by ESA Member States, NASA, and Canada.
This research has made use of data and software provided by the High Energy Astrophysics Science Archive Research Center (HEASARC), which is a service of the Astrophysics Science Division at NASA/GSFC and the High Energy Astrophysics Division of the Smithsonian Astrophysical Observatory.
This research has made use of the NASA/IPAC Extragalactic Database (NED),
which is operated by the Jet Propulsion Laboratory, California Institute of Technology,
under contract with NASA.

\newpage

\bibliography{jnb_sz_refs}{}
\bibliographystyle{aasjournal}



\end{document}